\documentclass[
  preprint,
]{revtex4-1}

\makeatletter
\newcommand{\whencolumns}[2]{\preprintsty@sw{#1}{#2}}
\makeatother

\usepackage[english]{babel}
\usepackage{amsmath,amssymb,amsfonts} 	
\usepackage{graphicx}
\usepackage{array,multirow}
\usepackage[utf8]{inputenc}
\usepackage{bm}
\usepackage{xcolor}
\usepackage[normalem]{ulem}
\usepackage{siunitx}
\usepackage{hyperref}
\hypersetup{colorlinks=true,linkcolor=blue,citecolor=blue,urlcolor=blue}
\usepackage{url}
\usepackage{textcomp}
\usepackage{lineno,hyperref}
\usepackage{float}
\usepackage{textcomp} 
\usepackage{rotating}
\usepackage[figurename=FIG.]{caption}
\usepackage{subcaption}
\usepackage{color} 
\definecolor{red}{rgb}{1.,0.,0.}
\usepackage{soul}
\usepackage{amsmath}

 \usepackage[
 justification=raggedright]
 {caption}
\usepackage{subcaption}

\makeatletter
\newcommand*{\rom}[1]{\expandafter\@slowromancap\romannumeral #1@}
\makeatother

\newcommand\THEOS{Theory and Simulation of Materials (THEOS)}
\newcommand\MARVEL{National Centre for Computational Design and Discovery of Novel Materials (MARVEL), {\'E}cole Polytechnique F{\'e}d{\'e}rale de Lausanne, 1015 Lausanne, Switzerland}
\newcommand\MARVELL{National Centre for Computational Design and Discovery of Novel Materials (MARVEL), {\'E}cole Polytechnique F{\'e}d{\'e}rale de Lausanne, 1015 Lausanne, Switzerland}
\newcommand\KINGS{Physics Department, King's College, London WC2R 2LS, United Kingdom}

\begin{document}

\title{Thermomechanical properties of honeycomb lattices from internal-coordinates potentials: the case of graphene and hexagonal boron nitride}
\author{Francesco Libbi}
\affiliation{\THEOS}\affiliation{\MARVEL}
\author{Nicola Bonini}
\affiliation{\KINGS}
\author{Nicola Marzari}
\affiliation{\THEOS}\affiliation{\MARVELL}

\begin{abstract}
Lattice dynamics in low-dimensional materials and, in particular, the
quadratic behaviour of the flexural acoustic modes play a fundamental
role in their thermomechanical properties. A first-principles evaluation
of these can be very demanding, and can be affected by numerical noise
that breaks translational or rotational invariance. In order to overcome
these challenges, we study the Gartstein internal-coordinate potential
and tune its 13 parameters on the first-principles interatomic force
constants for graphene. We show that the resulting potential not only
reproduces very well the phonon dispersions of graphene, but also those
of carbon nanotubes of any diameter and chirality. The addition of a
cubic term allows also to reproduce the dominant anharmonic terms,
leading to a very good estimate of the lattice thermal conductivity.
Finally, this potential form works very well also for boron nitride,
provided it is fitted on the short-range (analytical) part of the
interatomic force constants, and augmented thereafter with the
long-range dielectric contribution. This consideration underscores how
potentials based on short-ranged descriptors should be fit, in polar
materials, to the short-range part of the first-principles interactions,
and complemented by long-range analytical dielectric models
parametrized on the same first-principles calculations.

\end{abstract}

\maketitle

\section{Introduction}
\noindent
In the last decade, a sustained effort has focused on a number of low-dimensional materials \cite{Mounet2018,AKINWANDE201742} that exhibit outstanding properties and exotic phenomena, with promising applications for next-generation electronic and opto-electronic applications \cite{Radisavljevic2011,doi:10.1021/nn400280c}. The most studied case is undoubtedly graphene. Its thermal conductivity is among the highest measured \cite{Ghosh_2009,doi:10.1021/nl0731872}, and it is complemented by very large mechanical strength  \cite{PAPAGEORGIOU201775} and electronic mobility \cite{BOLOTIN2008351}. Moreover, carbon nanotubes are also intensely studied for their electronic, elastic and thermal properties \cite{Salvetat1999,PhysRevLett.118.135901}.
In order to characterize many of these properties, it is fundamental to accurately describe  their lattice dynamics. Here, the goal is to develop an approach that is computationally inexpensive but accurate enough to predict the potential energy of the lattice with respect to the atomic displacements up to the third derivatives; these latter determine phonon-phonon interactions, which control the dissipation of heat flux and the thermal conductivity, while second derivatives determine phonon dispersions and thermomechianical properties. \ 

We start from the internal-coordinate potential (ICP) introduced by Gartstein \cite{GARTSTEIN200483}, and tune it on first-principles calculations of interatomic force constants (IFCs) of either graphene or hexagonal (monolayer) boron nitride.
We show that the resulting ICPs reproduce very well harmonic properties, and in particular the quadratic behaviour of the flexural modes since they satisfy by construction translational and rotational invariance.
Such behaviour is a crucial feature in low-dimensional materials, greatly affecting the thermomechanical properties of these systems \cite{doi:10.1080/21663831.2016.1174163,doi:10.1021/nl202694m}.
Moreover, we show that such ICPs can be used for different purposes: they can be used to calculate phonons for nanotubes of any diameter and chirality, extended to other honeycomb based 2D materials, or augmented with anharmonic terms to describe thermal conductivities and thermal expansions. 
The open-source codes to calculate phonon dispersions and forces on atoms,  for the case of graphene, boron nitride and carbon nanotubes are provided in the Archive section of the Materials Cloud \cite{talirz2020materials,materialscloudentry}. Finally, it must be pointed out that the ICP method is general, and could be applied without restrictions to materials of any dimensionality.
\\

\section{The internal-coordinates potential}
\noindent

\begin{figure}[h]
\centering
\includegraphics[scale=0.27]{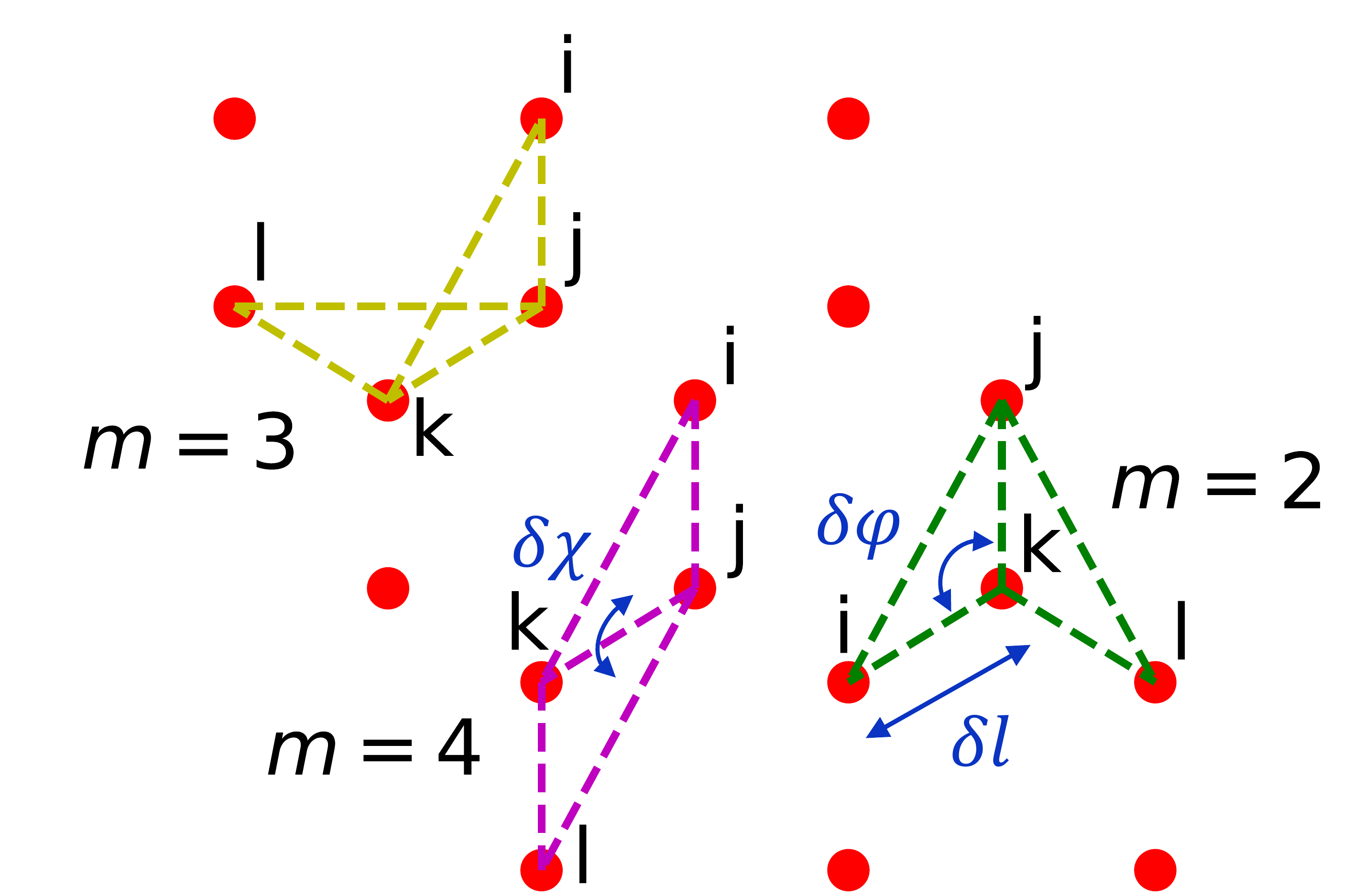}
\caption{There are three different kinds of dihedrals, labelled by the number $m$, which expresses the order of neighborhoodness between atoms $i$ and $l$ of the dihedral: e.g. atoms $i$ and $l$ in the green dihedral are second-nearest neighbours, so $\mathrm{\langle il \rangle =2}$. Following the same rule, $\mathrm{\langle il \rangle=3}$ for the yellow dihedral and $\mathrm{\langle il \rangle=4}$ for the violet one.}
\label{all_dihedrals}
\end{figure}

The analytic expression for the ICP, as originally introduced by Gartstein \cite{GARTSTEIN200483} for graphene, is given by the sum of the following two terms: \\

\begin{equation}\label{potential:1}
\begin{aligned}
\mathrm{
U_{int}=\sum_{\langle ijk\rangle} [K_1(\delta l_{ij}^2+\delta l_{jk}^2)+K_2\delta\varphi_{ijk}^2+
K_3\delta l_{ij}\delta l_{jk}+K_4\delta\varphi_{ijk}(\delta l_{ij}+\delta l_{jk})]} \\
\mathrm{
\,\,\,+\sum_{m=3}^4\sum_{\langle ijkl\rangle}^{\{il\}=m} [K_5^m\delta l_{ij}\delta l_{jk}+
K_6^m\delta\varphi_{ijk}\delta\varphi_{jkl}+ 
K_7^m(\delta\varphi_{ijk}\delta l_{kl}+\delta\varphi_{jkl}\delta l_{ij})]} \,,
\end{aligned}
\end{equation}
and
\begin{equation} \label{potential:3}
\mathrm{
U_{out}=\sum_{m=2}^4\sum_{ijkl}^{\{il\}=m}K_8^m\delta\chi_{ijkl}^2} \,.
\end{equation}
\\
The term $\mathrm{U_{int}}$ describes the in-plane deformation energy, while $\mathrm{U_{out}}$ describes the out-of-plane distortions.
To understand the meaning of the terms appearing in the ICP, it is necessary to group the atoms in triangular plaquettes $\mathrm{\langle ijk \rangle}$ formed by atoms linked by the bonds $\mathrm{\langle ij\rangle}$ and $\mathrm{\langle jk \rangle}$, and in dihedrals $\mathrm{ \langle ijkl \rangle}$ formed by two plaquettes $\mathrm{ \langle ijk\rangle}$ and $\mathrm{\langle jkl\rangle}$ sharing the bond $\mathrm{\langle jk\rangle}$. 
There are three different ways to form a dihedral (figure \ref{all_dihedrals}), labelled with the notation $\mathrm{\langle il\rangle =m}$,  which indicates that atoms $i$ and $l$ are the $m$-th nearest neighbours ($m$=2, 3, 4).\\
According to Eqs. \ref{potential:1}-\ref{potential:3}, the potential energy is a function of the variation of bond lengths $\mathrm{\delta l_{ij}}$, bond angles $\mathrm{\delta \phi_{ijk}}$ at atom $j$, and dihedral angles $\mathrm{\delta \chi_{ijkl}}$ between plaquettes belonging to the same dihedral.
Having a potential that depends on the variation of the internal coordinates offers two main advantages. The first is that the constraints on the IFCs due to translational and rotational invariance are automatically satisfied. This allows to reproduce correctly the acoustic branches near $\mathrm{\Gamma}$ without imposing any sum rule, resulting in quadratic dispersions for the flexural modes \cite{doi:10.1080/21663831.2016.1174163}. This feature is fundamental, since the dispersion of flexural mode and the scattering of acoustic phonons near the $\mathrm{\Gamma}$ point of the Brillouin zone \cite{doi:10.1021/nl202694m} are crucial to thermomechanical properties. The second advantage is that such ICPs can be applied to their respective nanotubes without any modification (neither a variation of the analytic expression nor a re-tuning of the parameters), since a nanotube can be obtained through an isometric mapping of the honeycomb sheet on a cylindrical surface, mantaining very good accuracy.\

The thirteen coefficients which parametrize the ICP above have been determined here by minimizing the mean square difference between the IFCs obtained from first-principles for graphene (see Section  \ref{methods} for details) and those determined through the ICP. Due to the large number of variables and local minima in the optimization problem, simulated annealing has been used with the purpose to obtain the global minimum independently from the initial guess. The set of parameters obtained are reported in the table \ref{k_params}.
\begin{table}
\caption{Values of the ICP parameters for graphene as optimized by a fit on the first-principles IFCs at the DFT-PBE level. The modulus of the Bravais vector adopted in the implementation of the ICP is 2.467 $\mathrm{\AA}$. The values of the parameters tuned on DFT-LDA calculations are provided in the Archive section of the Materials Cloud.} \cite{materialscloudentry} 
\label{k_params}
\begin{ruledtabular}
\renewcommand{\arraystretch}{1.2}
\begin{tabular}{lcr}
$\text{eV}\ \mathrm{\AA}^{-2}$ & $\text{eV}\ \text{rad}^{-2}$ & $\text{eV}\ \mathrm{\AA^{-1}}\ \text{rad}^{-1}$\\
\hline
$K_1=4.6611$ & $K_2=2.1612$ & $K_4=2.2254$  \\ 
$K_3=2.8943$ & $K_6^3=0.2412$ & $K_7^3=-0.1351$  \\ 
$K_5^3=-1.2939$ & $K_6^4=1.0493$ & $K_7^4=0.3034$  \\ 
$K_5^3=-0.5043$ & $K_8^2=0.1261$ & \multicolumn{1}{c}{} \\
\multicolumn{1}{c}{}  & $K_8^3=0.1110$ & \multicolumn{1}{c}{} \\
\multicolumn{1}{c}{}  & $K_8^4=0.2352$ & \multicolumn{1}{c}{} \\
\end{tabular}
\end{ruledtabular}
\end{table}\\
\begin{figure}[h]
\centering
\includegraphics[scale=0.32]{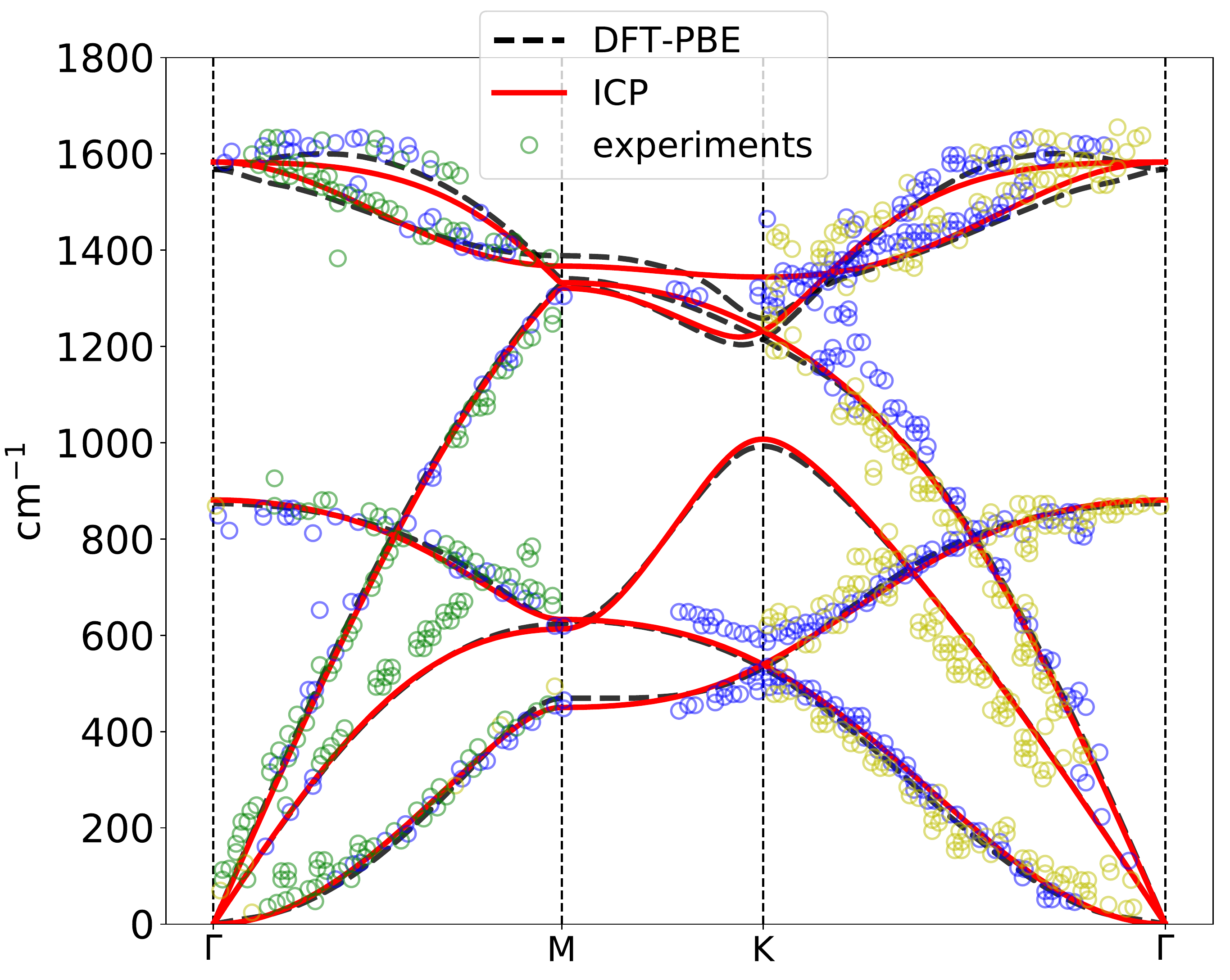}
\caption{Comparison between the results obtained using the ICP and those obtained from first-principles (DFT using the PBE exchange-correlation), for the case of graphene. The green, blue and yellow circles in the figure correspond to experimental values obtained using different techniques \cite{WIRTZ2004141}.}
\label{freqs}
\end{figure}
As it can be seen from figure \ref{freqs}, we find an excellent overall agreement between phonon frequencies as obtained from density functional theory using the PBE exchange-correlation functional and those obtained from the ICP; especially this is true for the acoustic branches, which greatly affect thermal conductivity at room temperature. The only difference between the two approaches is the absence in the ICP of the Kohn anomalies affecting the TO modes close to $\mathrm{\Gamma}$ and K \cite{PhysRevLett.93.185503}. However, since phonon branches around $\text{1500}\ \text{cm}^{-1}$ are almost completely empty at $\text{300}\,\text{K}$, this will not affect the relevant thermomechanical properties. We also remark that figure \ref{freqs} shows also the excellent agreement between both DFT and ICP results and experimental data, also compared to other potentials in the literature, as showed in Figs. 3 and 5 of Ref. \cite{PhysRevB.97.054303}. \

\begin{figure}[h]
\begin{subfigure}{0.40\textwidth}
\includegraphics[scale = 0.25]{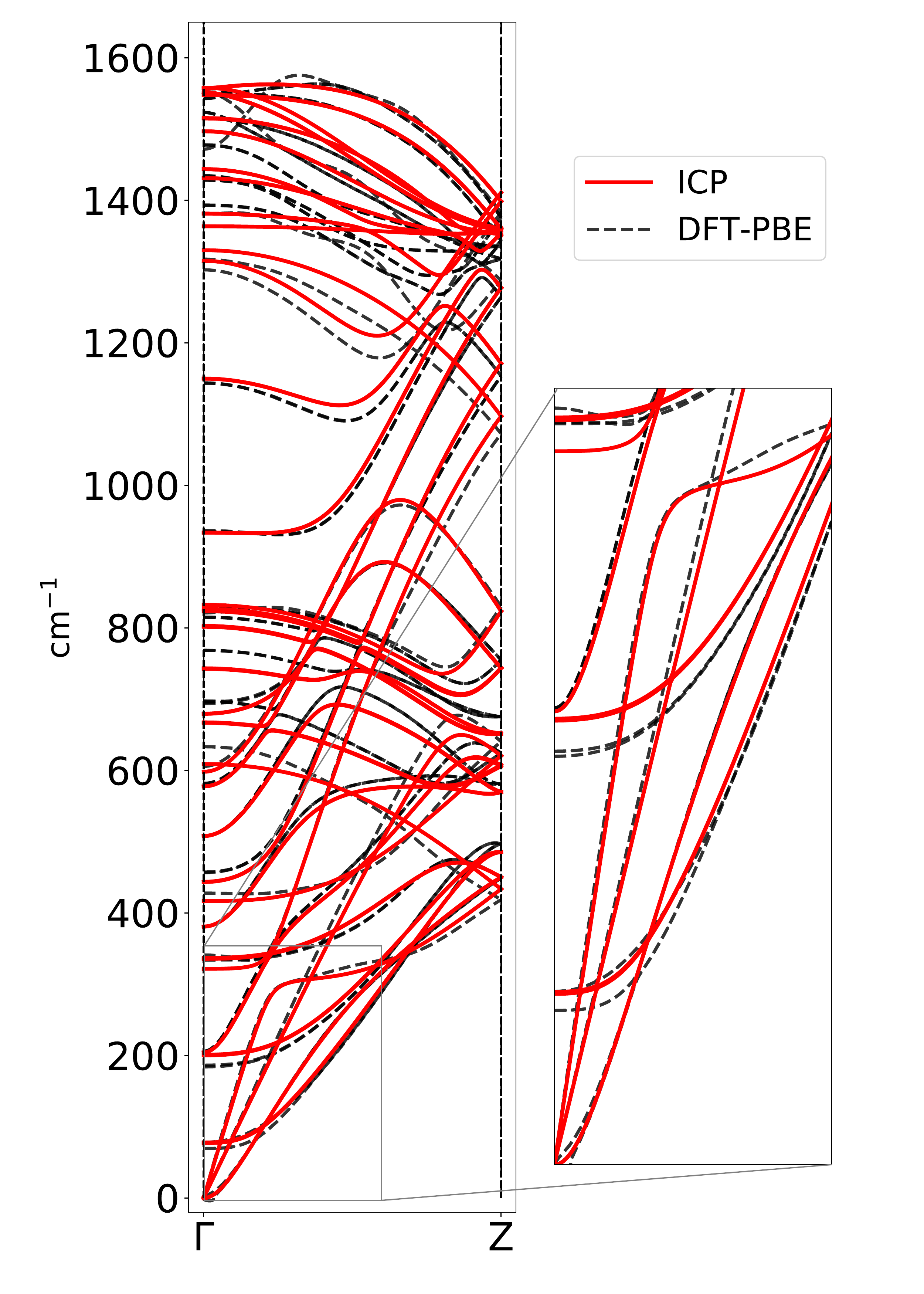}
\caption{(5,5) carbon nanotube}
\label {5_5_global}
\end{subfigure}
\begin{subfigure}{0.40\textwidth}
\includegraphics[scale = 0.25]{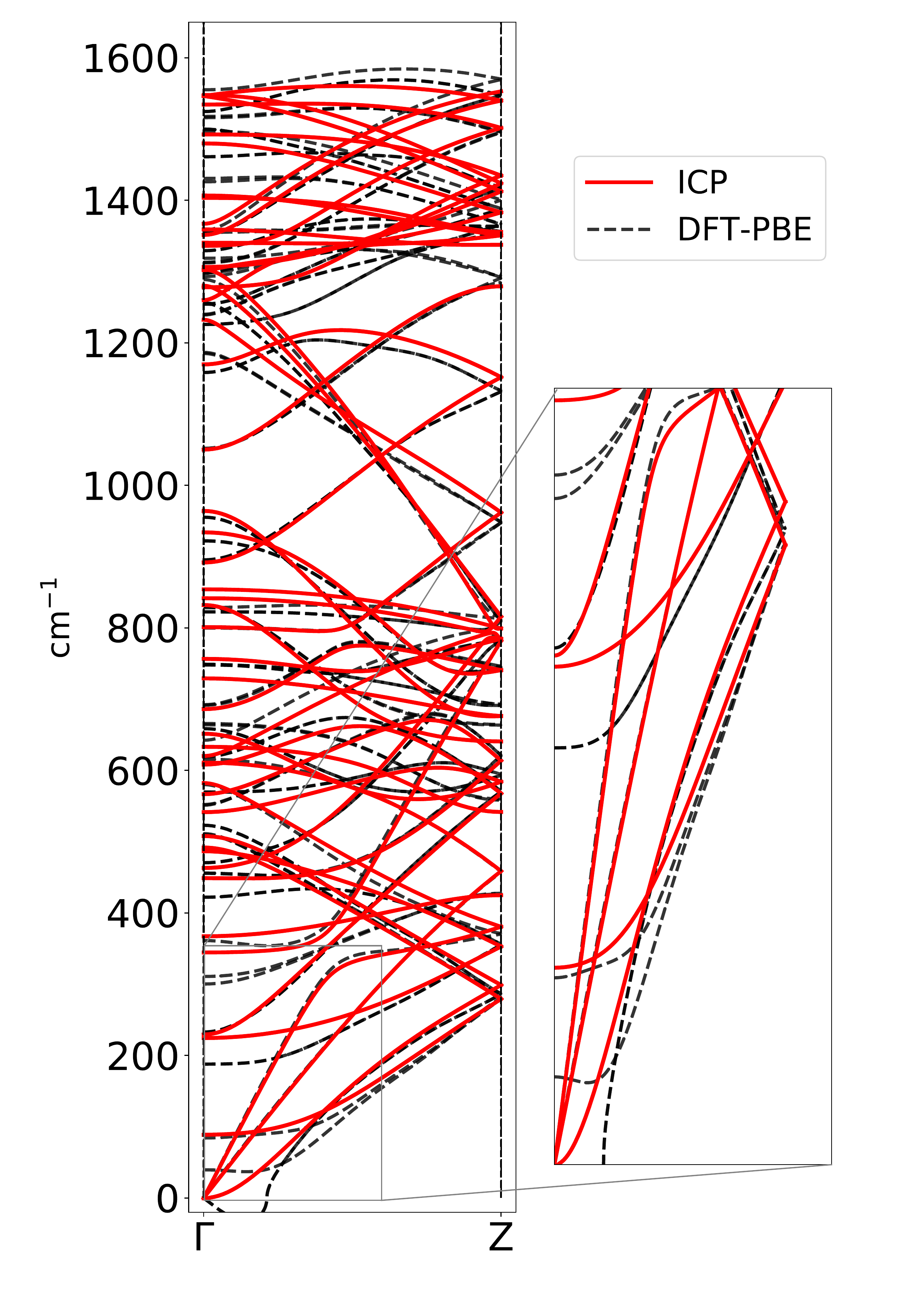}
\caption{(8,0) carbon nanotube}
\label {8_0_global}
\end{subfigure}
\caption{Panel (a) and panel (b) show, respectively,  the phonon dispersions for a (5,5) zig-zag carbon nanotube and for a (8,0) armchair carbon nanotube. The curves obtained through the ICP are shown as red continuous lines, while those obtained from first-principles (DFT-PBE) as black dashed lines.}
\label{cnt}
\end{figure}
\begin{figure}[h]
\includegraphics[scale=0.42]{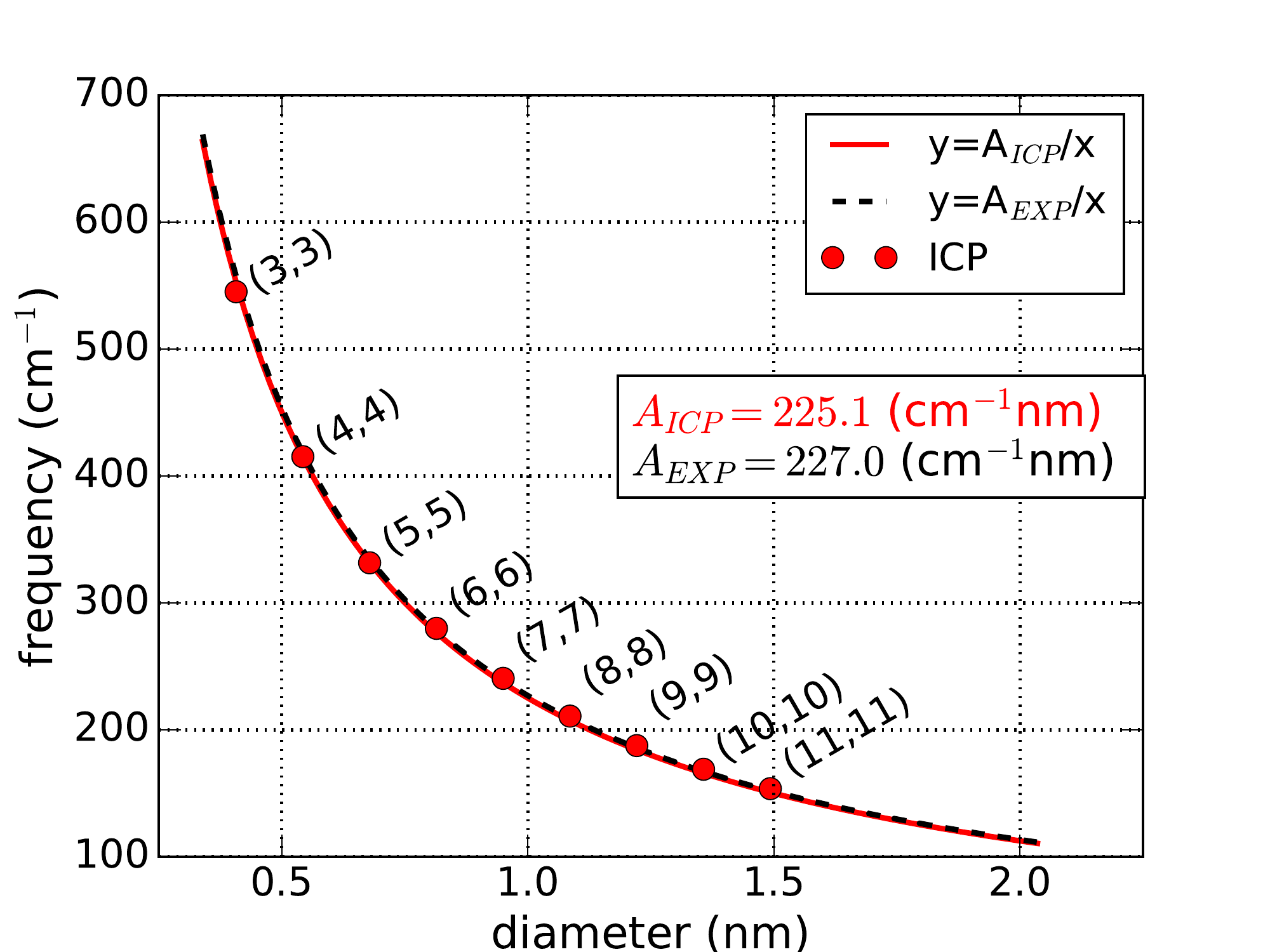}
\caption{Radial breathing mode for zig-zag carbon nanotubes as function of diameter. The theoretical values align perfectly on an hyperbole of equation $\text{y}=\text{A}/\text{x}$, with $\text{A}=\mathrm{225.1\ nm\ cm^{-1}}$. 
The experimental value \cite{PhysRevB.77.241403} for $\mathrm{A}$ is 227.0 nm $\mathrm{cm^{-1}}$, which is in excellent agreement with the theoretical prediction.}
\label{rbm}
\end{figure}

 
Furthermore, the resulting ICP can be used to compute phonons in carbon nanotubes of any diameter and chirality, without changing the value of the parameters used, by mapping the geometry of graphene on a cylindrical surface.
The phonon dispersion curves thus obtained (see figure \ref{cnt}) are in good agreement with those from DFT-PBE, especially in the lower part of the spectrum. As for the case of graphene, the main difference in the phonon spectra is the lack of Kohn anomalies \cite{PhysRevB.75.035427} \cite{PhysRevLett.88.235506}. It is also worth pointing out that first-principles modes close to $\mathrm{\Gamma}$ can sometimes display imaginary frequencies (shown as negative frequencies in the plot) as a result of Fourier transforms of the IFCs that do not satisfy acoustic sum rules, especially for rotations around an axis normal to the carbon nanotube (figure \ref{8_0_global} or Ref.  \cite{doi:10.1063/1.2753095}). There is clearly no trace of this unphysical result in the present ICP results.
Finally, it is interesting to show some comparison with experimental data also for carbon nanotubes. For this purpose, the radial breathing mode has been calculated for different zig-zag nanotubes, and the values obtained have been fitted with an hyperbole of equation $\text{y}=\text{A}/\text{x}$ (figure \ref{rbm}), obtaining for A a value of $\mathrm{225.1\ nm\ cm^{-1}}$, which is very close to the experimental value $\text{A}=\mathrm{227.0\ nm\ cm^{-1}}$ \cite{PhysRevB.77.241403}. \ 
The great accuracy of the ICP is achieved at a computational cost that is negligible, particularly when compared to first-principles calculations: the cpu-time required to perform a well converged phonon calculation for the (8,0) carbon nanotube on 6 points of the Brillouin zone is around $\mathrm{2\ 400}$ hours (8.6 Ms), against the 5 seconds of the ICP. The speedup is thus of the order of $\mathrm{10^6}$. 

The ICP can be used not only to compute phonons in carbon allotropes, but also to study the vibrational properties of all the materials characterised by an hexagonal lattice. Here we look next at a boron-nitride monolayer. As for the case of graphene, the ICP is tuned on first-principles IFCs for hexagonal boron nitride; since boron nitride contains two different kinds of atoms, the results can be improved by choosing different values for the parameter $K_2$  whether a boron atom ($K_2^B$) or a nitrogen atom ($K_2^N$) sit at the vertex of the angle $\phi_{ijk}$, and for the constant $K_8^2$ whether the dihedral $\langle ijkl\rangle $ is made up of one boron and three nitrogens ($K_8^{2,B}$) or one nitrogen and three borons ($K_8^{2,N}$), leading to 15 parameters that we tune using the same procedure outlined above.
\begin{figure}[h]
\centering
\includegraphics[scale=0.32]{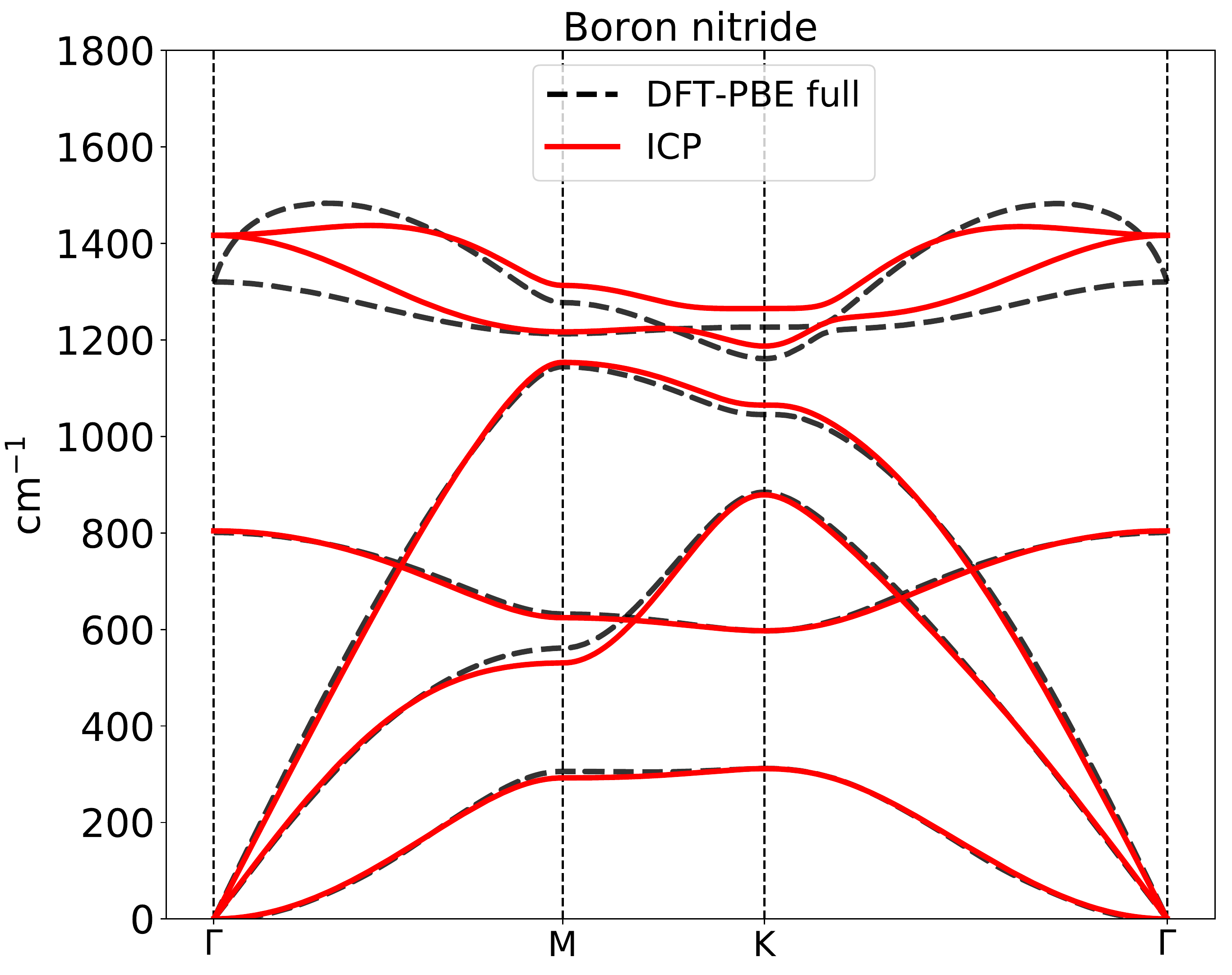}
\caption{Comparison between the phonon frequencies obtained using the ICP and those obtained from first-principles at the DFT-PBE level for hexagonal boron nitride, when the parameters of the ICP are tuned (incorrectly) to the  first-principles IFCs, which include at $\mathrm{q\neq0}$ long-range non-analytic terms. LO and TO modes in the DFT-PBE dispersions are degenerate at $\mathrm{\Gamma}$, but their slopes are different; as shown by Sohier et al.  \cite{PhysRevB.94.085415,doi:10.1021/acs.nanolett.7b01090}, this is a general feature for 2D polar materials.}
\label{bn}
\end{figure}
We show the resulting phonon dispersion in figure 5; while the agreement for the acoustic branches and the ZO mode is excellent, the mismatch for the LO and TO modes is due to strong long-range dielectric interactions \cite{PhysRevB.43.7231,RevModPhys.73.515} that are caused by the polarity of the material. In order to overcome this, one should tune the ICP parameters (table \ref{k_params_bn}) on the analytical part of the IFCs as obtained from first-principles after having subtracted at all $\mathrm{q\neq0}$ the non-analytic corrections (NACs) to the dynamical matrix \cite{PhysRevB.94.085415,doi:10.1021/acs.nanolett.7b01090}:
\begin{equation}
    \mathrm{
   D_{ai,a'j}(\mathbf{q}) = \frac{e^2}{\Omega} W_c(\mathbf{q}_p) 
   \frac{(\mathbf{q}_p\cdot \mathbf{Z}^*_a)_i (\mathbf{q}_p\cdot \mathbf{Z}^*_{a'})_j}{\sqrt{M_a M_{a'}}} }\ ,
\end{equation}
where $\Omega$ is the volume of the unit cell, $\mathbf{Z}^*_a$ is the Born effective charge tensor of the atom $a$ in the unit cell and $\mathrm{W_c(\mathbf{q}_p)}$ is the screened Coulomb interaction, which for 2D monolayers reads
\begin{equation}
   \mathrm{
    W_c(\mathbf{q}_p) = \frac{2\pi}{|\mathbf{q}_p| \Bigl( \epsilon_{0} + \frac{\mathbf{q}_p\cdot \mathbf{r}_{eff} \cdot \mathbf{q}_p }{|\mathbf{q}_p|^2} |\mathbf{q}_p|\Bigr)} }\ .
\end{equation}
These NACs can be then added on top of the IFCs generated through the ICP and fitted on the analytical part of the first-principles calculations.\
\begin{figure}
    \begin{subfigure}{0.5\textwidth}
    \includegraphics[scale = 0.32]{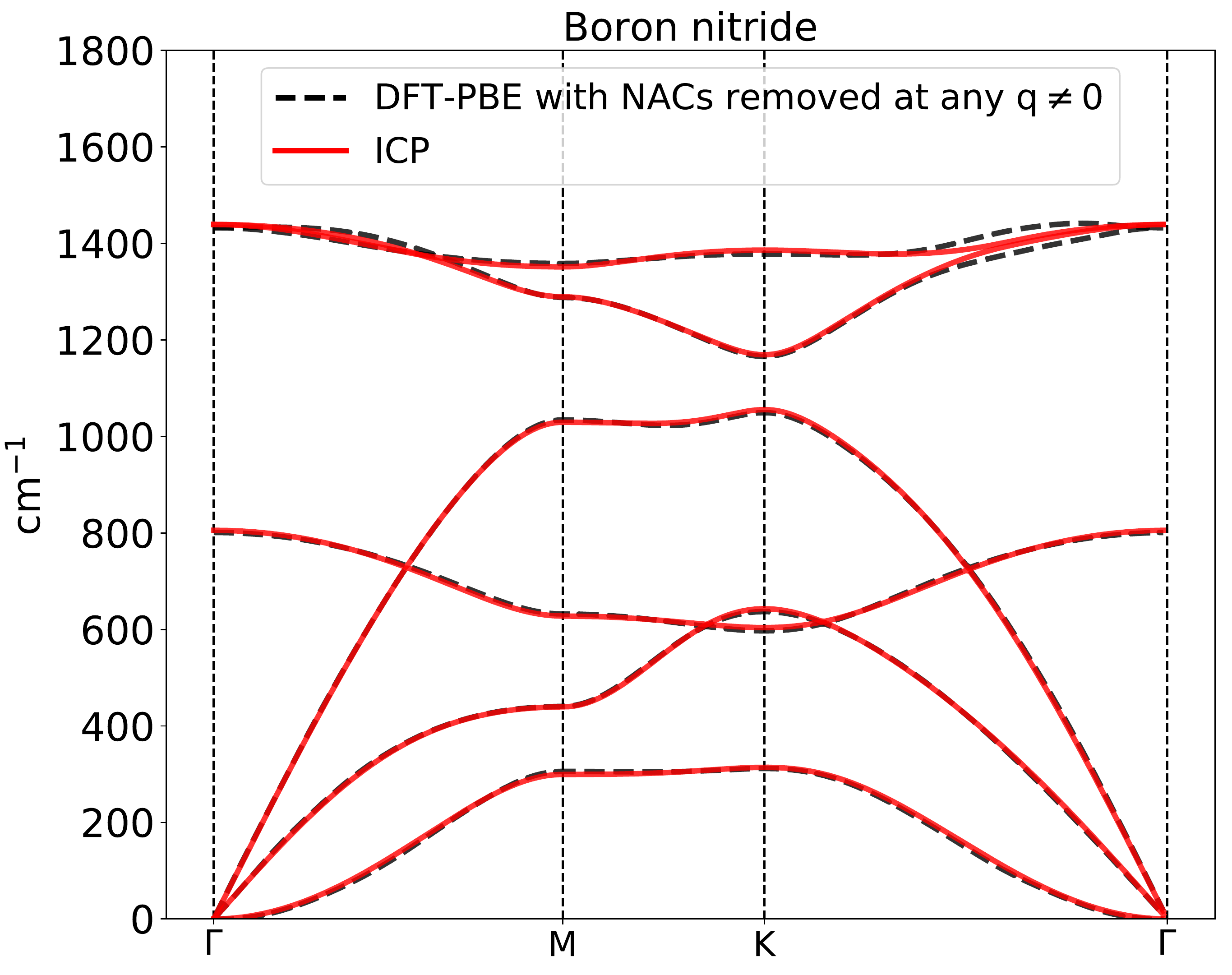}
    \caption{NACs removed at any $\mathrm{q\neq0}$.}
    \end{subfigure}
    \begin{subfigure}{0.5\textwidth}
    \includegraphics[scale = 0.32]{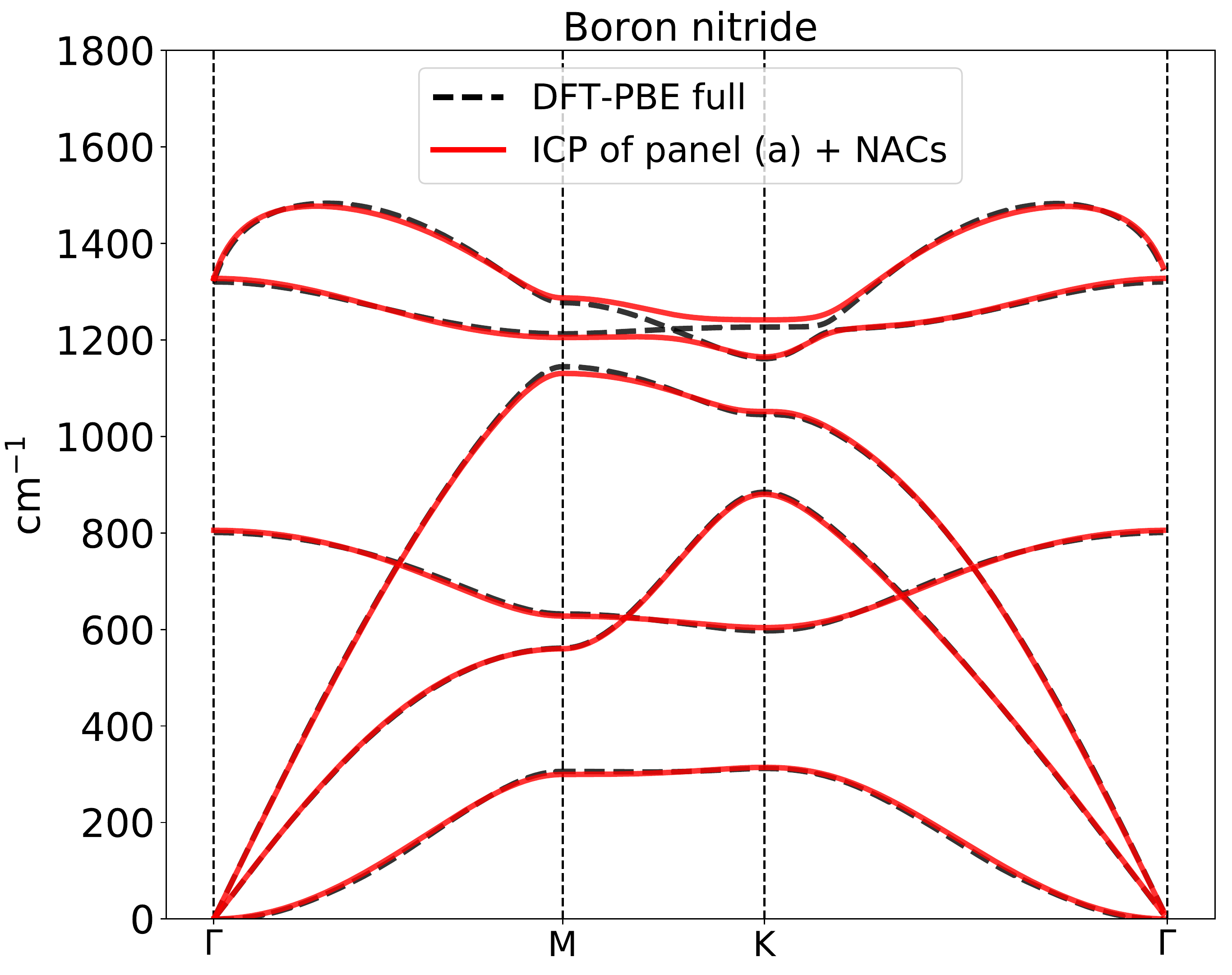}
    \caption{NACs added back at any q.}
    \end{subfigure}
    \caption{When the parameters of the ICP are tuned on first-principles IFCs with NACs removed at any $\mathrm{q\neq0}$, the agreement is almost perfect, as shown in panel (a).  Finally, when applying the NACs to both the ICP and the first-principles results, a very close matching is obtained also for the optical phonons (panel (b) ); this is the final result for monolayer hBN. }
    \label{analytic}
\end{figure} 
In order to perform the step mentioned above it is necessary to know the Bravais vectors, the high-frequency limit of the dielectric tensor $\bm{\mathrm{\varepsilon}_{\infty}}$ and the Born effective charges for boron ($\mathbf{Z}^{\star}_B$) and nitrogen ($\mathbf{Z}^{\star}_N$); the values used here are those determined from DFT-PBE:
\begin{equation}\label{eps-Z}
\begin{aligned}
|\mathbf{a}|= 2.501 \mathrm{\AA}\ , \qquad\qquad\qquad \\
\bm{\mathrm{\varepsilon}_{\infty}}= \bm{\mathrm{1}} - \frac{1}{\mathrm{V}}\, \mathrm{diag}(\,457.96,\, 457.96,\, 0.00\,)\ ,\\ 
\mathbf{Z}^{\star}_B= \mathrm{diag}(\,2.7267,\, 2.7267,\, 0.0000\,)\ ,\;\, \\ 
\mathbf{Z}^{\star}_N = -\mathbf{Z}^{\star}_B\ , \qquad\qquad\qquad
\end{aligned}
\end{equation}
where $\bm{\mathrm{1}}$ is the identity matrix, and the volume is in atomic units. The presence of the volume in Eq. \ref{eps-Z} is aimed to remove the arbitrariness in the definition of the dielectric tensor, linked to the fact that, for 2D dimensional materials, the size of the cell in the out-of-plane direction is a free parameter which must be converged to eliminate spurious interactions between periodic images.
In order to better understand the dependence of Eq. \ref{eps-Z} on the volume, it is useful to consider the analytic expression of the dielectric tensor in the framework of density-functional perturbation theory (DFPT) \cite{RevModPhys.73.515}):
\begin{equation}\label{eps-dfpt}
    \mathrm{
    \varepsilon_{\alpha\beta}^{\infty}= \delta_{\alpha\beta} -\frac{1}{V} \, \Bigl( \frac{16 \pi e}{E_{\beta}} \sum_{n=1}^{N/2} \langle \bar{\psi}^{\alpha}_n |\Delta^{E_{\beta}} \psi_n\rangle \Bigr)} \ ,
\end{equation}
where $\mathrm{E_{\beta}}$ is the perturbing electric field and $\mathrm{|\Delta^{E_{\beta}} \psi_n\rangle}$ is the variation of the Kohn-Sham wave function $\mathrm{\psi}_n$ due to the perturbation. When increasing the size of the cell in the out-of-plane direction, the part enclosed in the round brackets in the rhs of Eq. \ref{eps-dfpt} converges toward a fixed value, while the volume grows linearly; therefore, different choices of the volume lead to different values for $\mathbf{\mathrm{\varepsilon}_{\infty}}$. This arbitrariness in the definition of the dielectric constant does not affect the NACs, provided that the volume used when applying them is equal to that adopted in the DFPT calculation of the dielectric tensor.
The out-of-plane component of both the dielectric tensor and the effective charges does not affect the NACs, therefore it has been set to zero in Eq. \ref{eps-Z}.
The results obtained by fitting the potential on the analytic part of the IFCs and adding the NACs is reported in figure \ref{analytic}, showing a perfect agreement between the ICP and first-principles predictions. These considerations are also very relevant for machine-learned potentials, where first-principles calculations are fitted with neural networks or kernel regressions methods on local representations \cite{PhysRevB.81.184107, PhysRevLett.104.136403,PhysRevB.87.184115}, suggesting that the fit could be performed on forces or IFCs purified from the long-range non-analytic behaviour at $\mathrm{q\neq0}$, while the non-analytic effects should be determined in reciprocal space and then summed back. In alternative, the full potential could be fitted by incorporating the non-local information within the machine-learning representation, following the work of Ref. \cite{doi:10.1063/1.5128375}.
We note in passing that for phonons in hBN nanotubes one would need NACs for one-dimensional systems \cite{not-published}.
\begin{table}[h]
\caption{Values of the ICP parameters for hexagonal boron nitride. The length of the Bravais vector used in the ICP is 2.501 $\mathrm{\AA}$ (obtained from DFT-PBE calculations). The parameters are tuned in order to reproduce correctly the IFCs calculated from first-principles with NACs subtracted at any $\mathrm{q\neq0}$ \cite{doi:10.1021/acs.nanolett.7b01090}. These NACs, with the dielectric tensor and effective charges as described in the text, are then added back to the ICP in reciprocal space.}
\label{k_params_bn}
\begin{ruledtabular}
\renewcommand{\arraystretch}{1.2}
\begin{tabular}{lcr}
$\text{eV}\ \mathrm{\AA}^{-2}$ & $\text{eV}\ \text{rad}^{-2}$ & $\text{eV}\ \mathrm{\AA^{-1}}\ \text{rad}^{-1}$\\
\hline
$K_1=4.4152$ & $K^B_2=1.1339,\,\, K^N_2=1.8419$ & $K_4=2.3921$  \\ 
$K_3=1.9252$ & $K_6^3=0.6647$ & $K_7^3=0.0944$  \\ 
$K_5^3=-0.8098$ & $K_6^4=0.5619$ & $K_7^4=0.1257$  \\ 
$K_5^3=-0.1941$ & $K_8^{2,B}=0.2725,\,\, K_8^{2,N}=-0.0016$ & \multicolumn{1}{c}{} \\
\multicolumn{1}{c}{}  & $K_8^3=0.0766$ & \multicolumn{1}{c}{} \\
\multicolumn{1}{c}{}  & $K_8^4=0.1480$ & \multicolumn{1}{c}{} \\
\end{tabular}
\end{ruledtabular}
\end{table}\\
\section{Second and third order interatomic force constants}
In order to widen the applicability of the ICP, one would like to reproduce not only the harmonic IFCs (i.e. the second derivatives of the potential energy with respect to the displacement of atoms in the supercell), but also the third-order IFCs. The former are directly related to phonons, as phonon dispersions are obtained by diagonalising the dynamical matrix -- which is the Fourier transform of the IFCs -- at any q vector in the Brillouin zone. The latter are instead related, in the language of second quantization, to the 3-body phonon-phonon interactions and determine the lifetimes that appear in the scattering term of the Boltzmann equation \cite{doi:10.1002/andp.19293950803} and control the heat flux dissipation.\ 

\begin{figure}[h]
\centering
\includegraphics[scale=0.43]{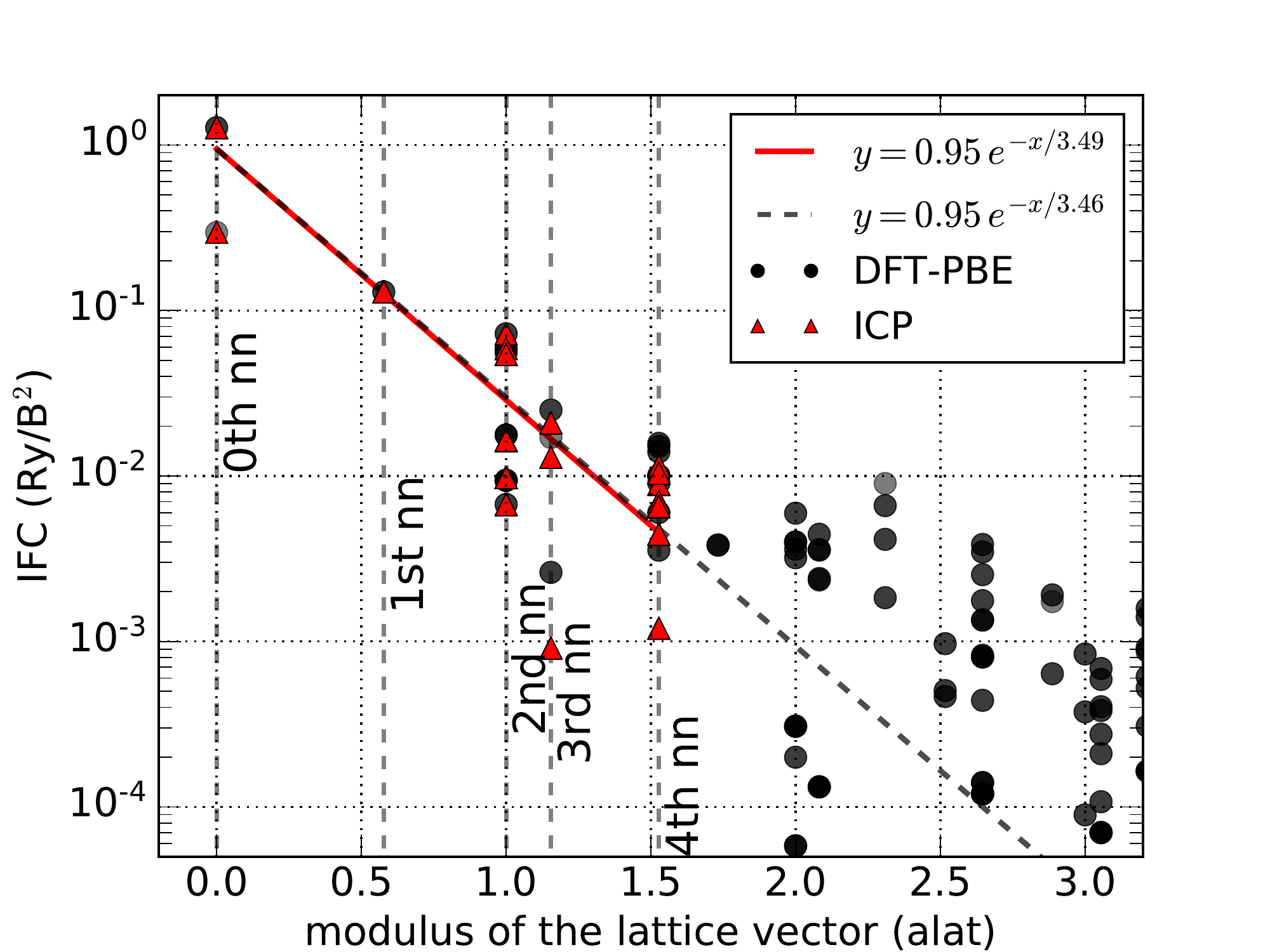}
\caption{IFCs for graphene as a function of the modulus of the lattice vector to which they correspond. The ICP takes into account interactions up to fourth nearest neighbours; therefore, the IFCs vanish for higher distances. A decreasing exponentials $A\,e^{-bx}$ is used to fit the behaviour of IFCs with the distance, both for the ICP (red solid line) and for the first-principles results (black dashed lines). The fitted parameters are almost equal in the two cases, another proof of the good match between second derivatives.}
\label{d2}
\end{figure}
 As shown in figure \ref{d2} for graphene, the IFCs calculated for the ICP are in excellent agreement with those obtained from first-principles.
The largest, short-ranged IFCs match perfectly, and even going to 4-th nearest neighbours the maximum difference between two corresponding IFCs is lower than $\text{1.5 10}^{-2}\ \text{Ry}\ \text{B}^{-2}$, which corresponds to 1.10$\%$ of the maximum IFC ($\text{1.36}\ \text{Ry}\ \text{B}^{-2}$, as calculated from first-principles). We note in passing that, up to the 4-th nearest neighbours, the IFCs sit on a decreasing exponential, while for larger interatomic distances the decay law changes due to the periodic-boundary conditions  in the calculation of IFCs for couples of atoms that are far from each other; a finer sampling of the dynamical matrix in reciprocal space would thus be required.\

\begin{figure}[h]
\begin{subfigure}{0.48\textwidth}
\includegraphics[scale = 0.36]{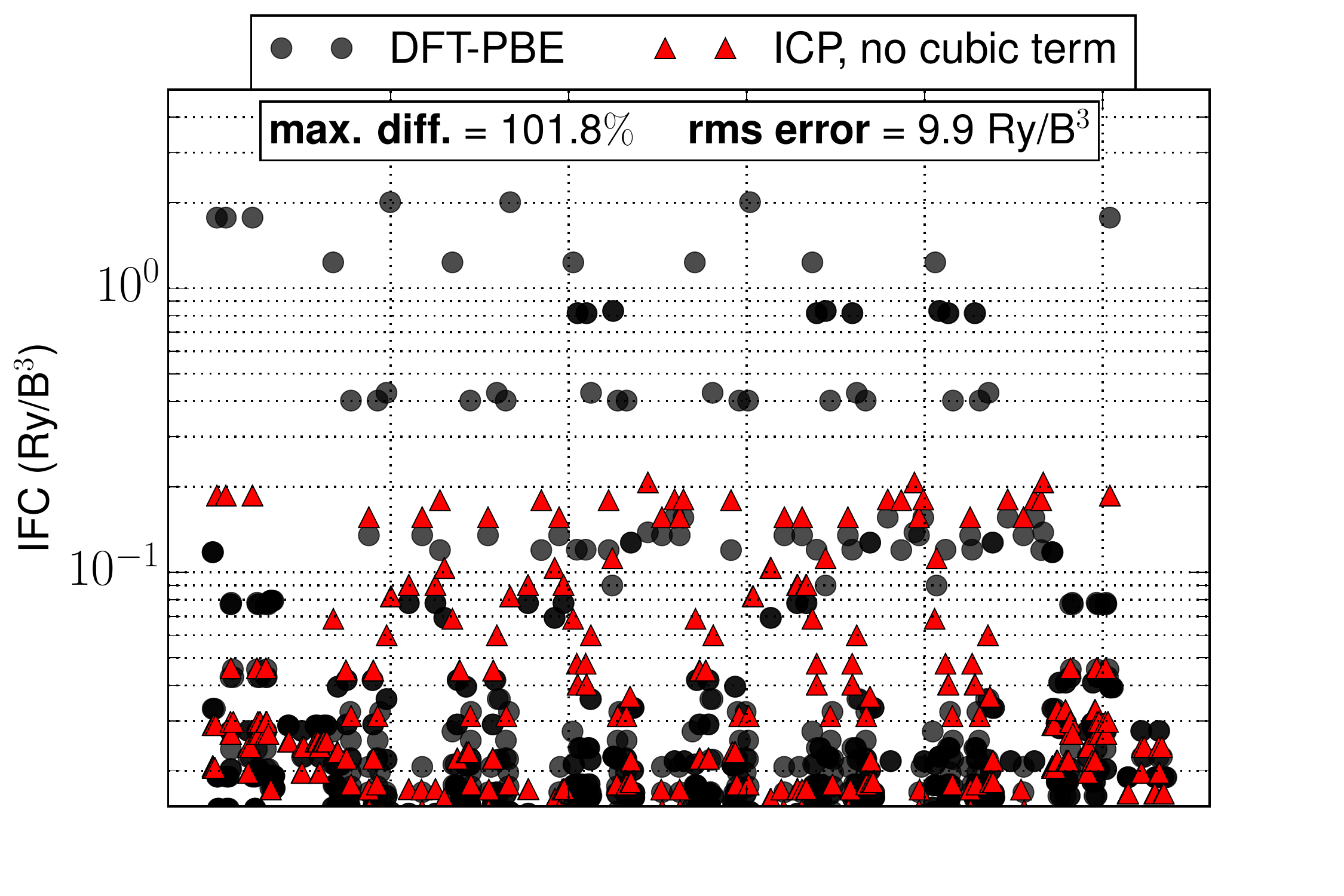}
\label{without}
\caption{Without cubic term.}
\end{subfigure}
\begin{subfigure}{0.48\textwidth}
\includegraphics[scale = 0.36]{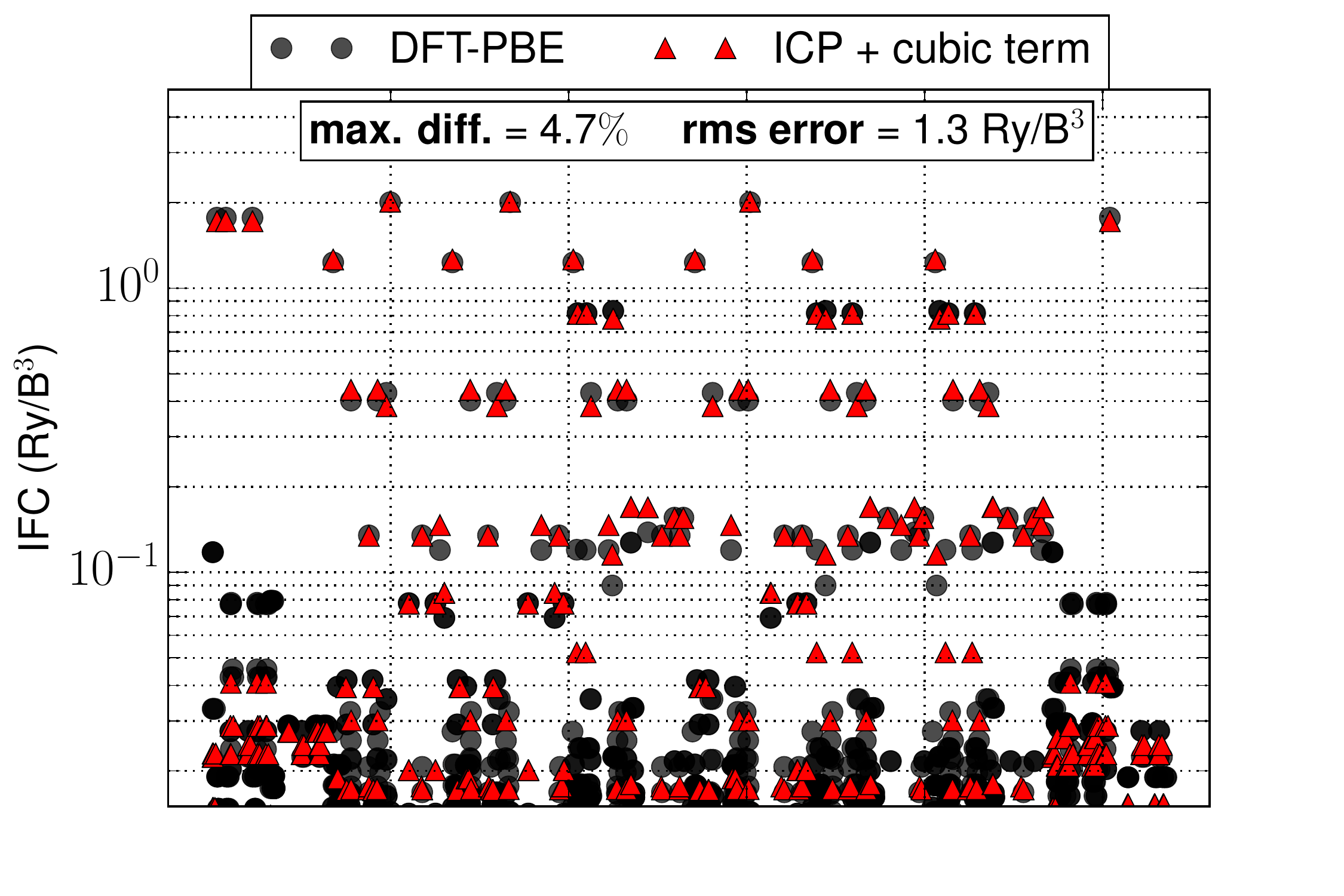}
\label{with}
\caption{Cubic term added.}
\end{subfigure}
\caption{Panel (a), comparison of the third derivatives obtained from first-principles for graphene with those calculated through the original ICP. It is easy to note that no red triangle is superposed to the black dots corresponding to the highest terms. When adding a single cubic term $\mathrm{K_{d3}\delta l^3}$ (panel (b) ), a good overall improvement is achieved, and all the leading terms match very well.}
\label{d3}
\end{figure}

Although the ICP (Eqs. \ref{potential:1}-\ref{potential:3}) contains some anharmonicity, it turns out to be negligible: the largest third derivative generated using the ICP is around $\text{0.2}\,\text{Ry}\ \text{B}^{-3}$, ten times smaller than the largest derivative calculated from first-principles, and overall there is a difference of one order of magnitude between the largest corresponding third derivatives (see figure \ref{d3}, panel (a) ).
However, the dominant anharmonic effects in the potential of graphene can be captured by adding a single extra term, in the form of a stretching cubic contribution $\text{K}_{\text{d3}}\,\mathrm{\delta}\text{l}^3$. It is important to stress that since a cubic term does not alter the second derivatives at the equilibrium configuration, there is no need to tune again the original ICP, whose anharmonicity is negligible.
Furthermore, since only the parameter $\text{K}_{\text{d3}}$ has to be determined, it is possible to do this easily, without using any minimization technique; we do this by imposing that the largest third derivative is reproduced correctly.
The panel (b) of figure \ref{d3} shows the result: with a value for $\text{K}_{\text{d3}}$ of $\text{-7.83}\ \text{Ry}\ \text{B}^{-3}$, the mean root square error decreases from $\text{10.3}\ \text{Ry}\ \text{B}^{-3}$ to $\text{1.3}\ \text{Ry}\ \text{B}^{-3}$, while the maximum difference between corresponding third derivatives goes from 101.7$\%$ of the largest first-principles derivative to only 4.7$\%$. 
\\

\section{Effect of strain}
\begin{figure}[h]
\begin{subfigure}{0.48\textwidth}
\includegraphics[scale = 0.27]{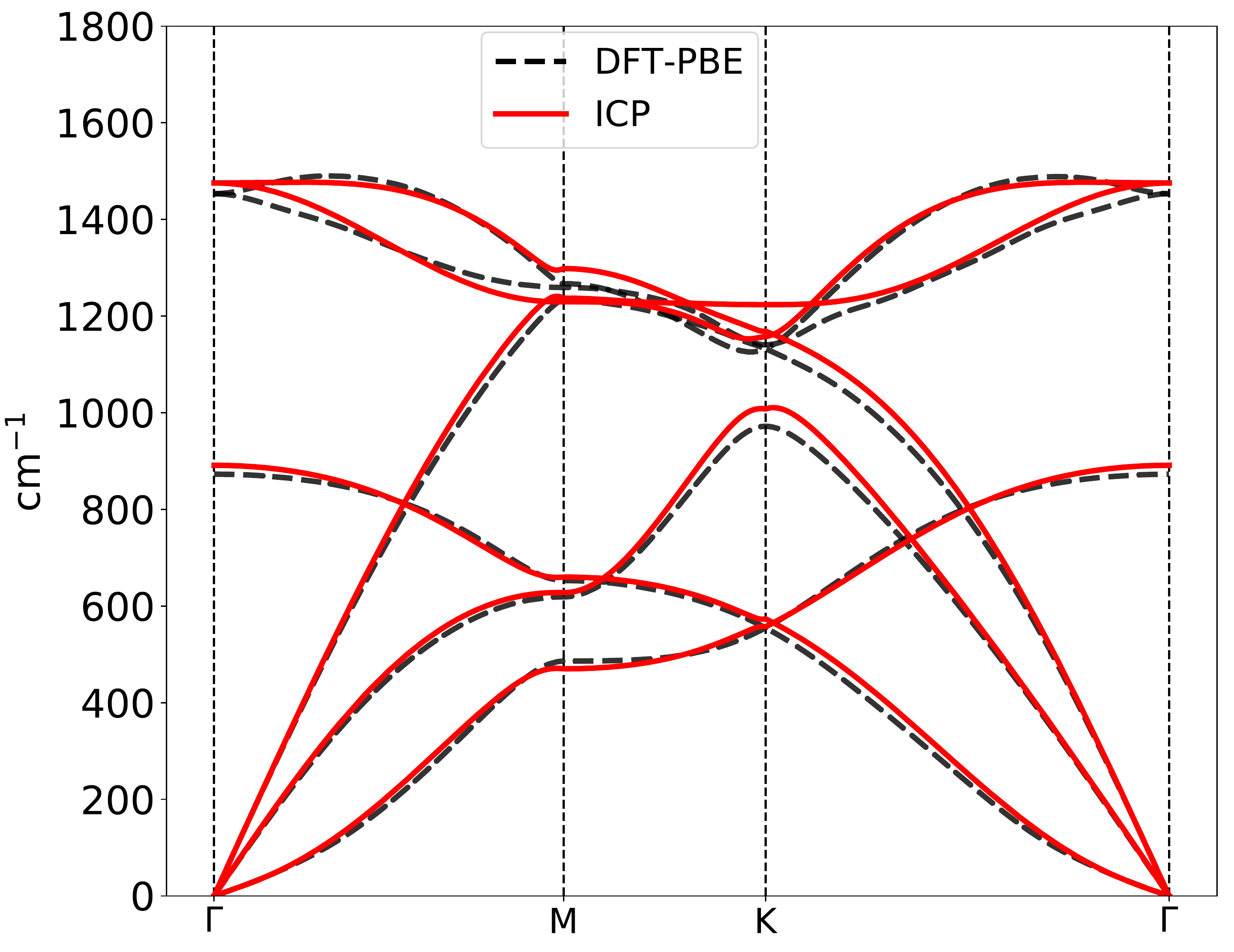}
\label{without}
\caption{2\% biaxial strain.}
\end{subfigure}
\begin{subfigure}{0.48\textwidth}
\includegraphics[scale = 0.27]{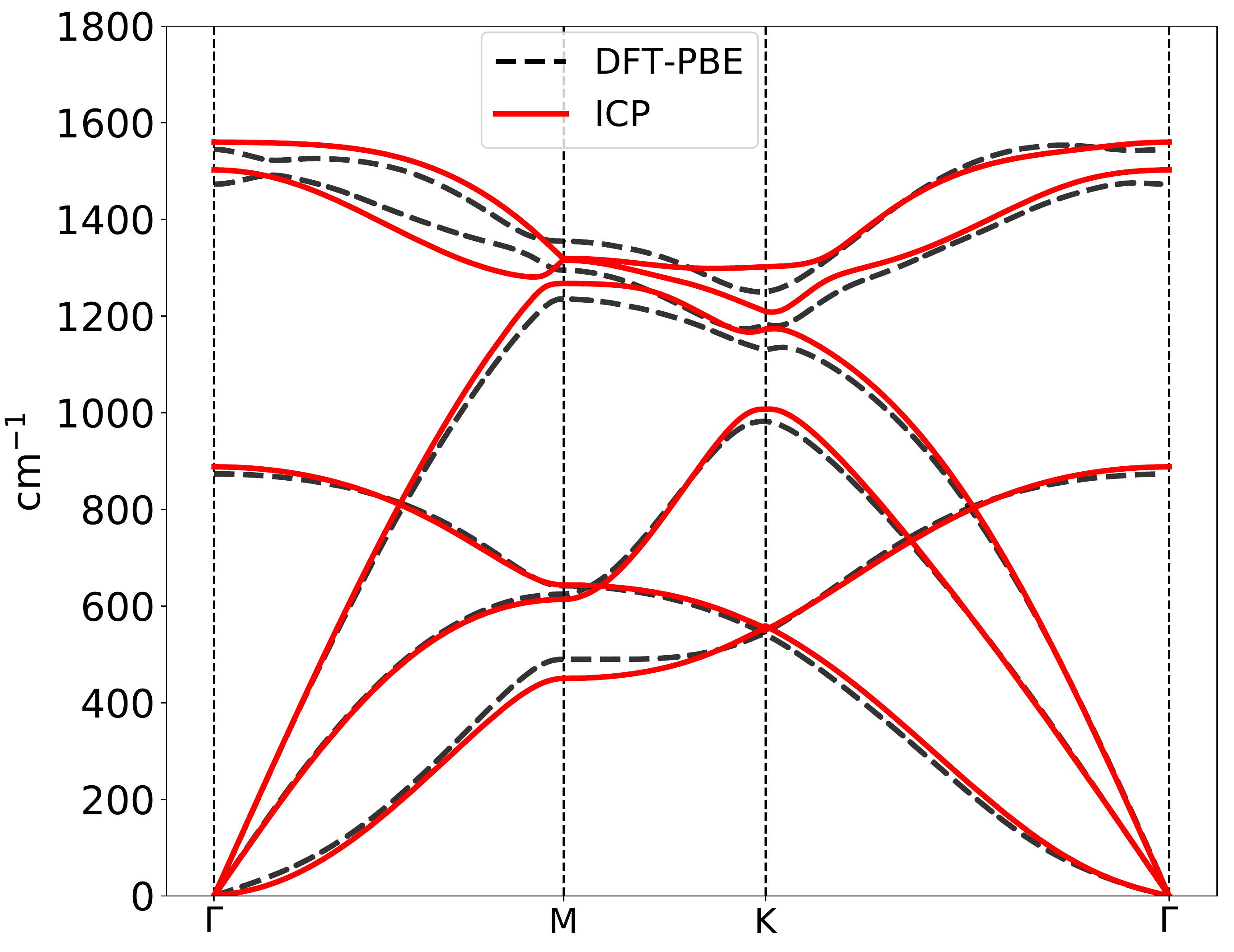}
\label{with}
\caption{2\% uniaxial strain in the zig-zag direction.}
\end{subfigure}
\caption{Panel (a) and (b) show the phonon the dispersions for the ICP and DFT-PBE obtained after imposing a 2\% biaxial and uniaxial strain, respectively.}
\label{strain}
\end{figure}

\begin{figure}[h]
\centering
\includegraphics[scale=0.7]{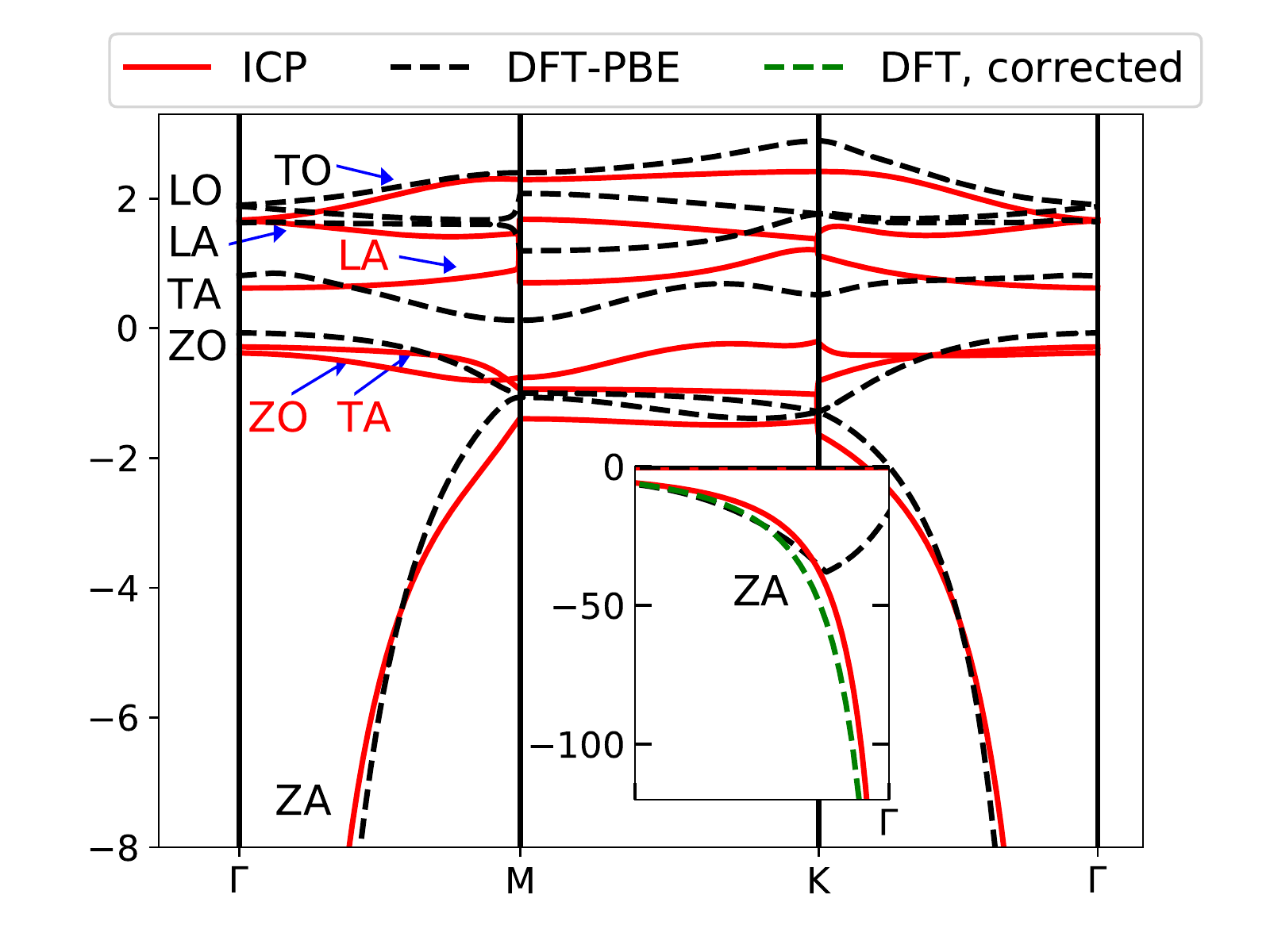}
\caption{Gruneisen parameters calculated from first-principles (black dashed line) and with the ICP (red solid line) along a high-symmetry path in the Brillouin zone. Moving from the bottom to the top (at $\mathrm{\Gamma}$), the branches correspond to ZA, ZO, TA, LA, LO, TO for both DFT-PBE and ICP . For clarity, selected labels for DFT-PBE (black) and ICP (red) branches are displayed. The inset shows the long-wavelength behavior of the ZA modes. The green dashed line represents the first-principles results after the ZA modes have been fitted to a parabola in the neighbourhood of $\mathrm{\Gamma}$, as explained in the main text. }
\label{gruneisen}
\end{figure}

The augmentation of the ICP with the cubic term is essential also for reproducing phonons in a strained geometry. Phonons in graphene for 2\% biaxial and uniaxial strains (in the zig-zag direction) are reported in figure \ref{strain}.
For the biaxial strain the agreement is excellent throughout all the frequency range. The uniaxial strain case shows some minor mismatch in the higher part of the spectrum, which, however, is not populated at room temperature. Since the application of a uniaxial strain breaks the hexagonal symmetry of the lattice, it is necessary to relax first the atomic positions. 
It is interesting to compare the predictions of the relative displacements of the atoms in the unit cell caused by the relaxation: the ones calculated from first-principles are  $0.0074\ \mathrm{a_0}$ in the zig-zag direction and $0.0016\ \mathrm{a_0}$ in the armchair direction, while those determined by the ICP are  $0.0077\ \mathrm{a_0}$ in the zig-zag direction and   $0.0018\ \mathrm{a_0}$ in the armchair direction ($\mathrm{a_0}$ is the unstrained lattice parameter), in very good agreement with the first-principles predictions.\\
In order to have another measure of the anharmonicity of the ICP, we calculate the Gr\"uneisen parameters, defined as in Ref. \cite{PhysRevB.71.205214}:
\begin{equation}\label{gru}
\mathrm{
\gamma_{\mathbf{q},s} = -\frac{1}{2\omega^0_{\mathbf{q},s}} \frac{d \omega_{\mathbf{q},s}}{d \epsilon} \Bigr |_0    }\ ,
\end{equation}
where $\mathrm{\omega^0_{\mathbf{q},s}}$ is the unstrained frequency and $\mathrm{\epsilon}$ the biaxial strain.
These are shown in figure \ref{gruneisen}.
A good match is found for the ZA, ZO, LO and TO modes (respectively, the first, second, fifth and sixth branches starting from the bottom), while the TA and LA parameters (third and fourth lines)  predicted by the ICP are, quite rigidly, down-shifted by $\sim$1. There is an apparent difference in the behavior of the parameters for the ZA mode in the long-wavelength limit. The curves obtained from first-principles and with the ICP remain superimposed until a certain wavelength; for larger wavelengths/shorter q wavevectors which the first-principles line reverts its trend, while the ICP line continues following its asymptotic behaviour, and diverges as $\mathrm{q^{-2}}$. The reason for this discrepancy lies in the fact that the first-principles parameters are affected by the incorrect prediction of the ZA frequencies very close to the $\mathrm{\Gamma}$ point. This incorrect behaviour is due to residual numeric noise, which is not removed by the acoustic sum rule, and causes the appearance of a spurious linear dependence of $\mathrm{\omega^0_{\mathbf{q},s}}$ on $\mathrm{|q|}$. Therefore, when dividing $\mathrm{\frac{d \omega_{\mathbf{q},s}}{d \epsilon}}$ by $\mathrm{\omega^0_{\mathbf{q},s}}$ and taking the limit $\mathrm{q\rightarrow 0}$, the spurious linear term dominates on the correct quadratic term, thus eliminating the divergence. We note also that the discussion on the quadraticity of the ZA mode in the long-wavelength limit is still open (see Ref. \cite{10.1103/PhysRevB.97.035426} for a discussion of the current state of the negative thermal expansion in graphene, and Ref. \cite{ADAMYAN20163732} discussing linearity in the long-wavelength limit), so we consider here the quadratic behaviour as a limiting case for an ideal membrane.
As mentioned, this happens only in a very small region around $\mathrm{\Gamma}$, outside of which the quadratic behaviour of the ZA mode is reproduced correctly by first-principles calculations. Therefore, it is easy to correct the error on the first-principles ZA Gr\"uneisen parameters by fitting the ZA frequencies with a parabola in the region where they are reproduced correctly and are quadratic ($\mathrm{0.033\frac{2\pi}{|a_0|} < |\mathbf{q}| < 0.267\frac{2\pi}{|a_0|}}$), and using this parabola in the noisy region of very short wave vectors ($\mathrm{|\mathbf{q}| < 0.033\frac{2\pi}{|a_0|}}$). The Gr\"uneisen ZA mode obtained using the fitted frequencies is represented with a green dashed line in the inset of figure \ref{gruneisen}, and is almost superimposed to the ICP prediction, which did not need a sum rule or a quadratic fit. This can be regarded as a further proof of the accuracy of the ICP in reproducing the acoustic modes.\\
The $\mathrm{q^{-2}}$ divergence of the ZA parameters has important implications \cite{PhysRevB.86.144103} in the evaluation of the linear thermal expansion coefficient, which is defined as $\mathrm{\alpha = \frac{1}{a_0} \frac{da}{dT} = \frac{d\epsilon}{dT}}$ and can be calculated starting from the Gr\"uneisen parameters $\mathrm{\gamma_{\mathbf{q}s}}$ under the framework of the quasi-harmonic approximation (QHA):
\begin{equation}\label{alpha} \mathrm{
\alpha = \frac{2}{\frac{d^2E}{d\epsilon^2}} \sum_{\mathbf{q},s} c_v(\mathbf{q},s) \gamma_{\mathbf{q},s} }\ ,
\end{equation}
where E represents the electronic energy of the crystal and $\mathrm{c_v(\mathbf{q},s)}$ the specific heat of a phonon mode of wavevector $\mathrm{\mathbf{q}}$ and branch index s.
\begin{figure}[h]
\begin{subfigure}{0.48\textwidth}
\includegraphics[scale = 0.5]{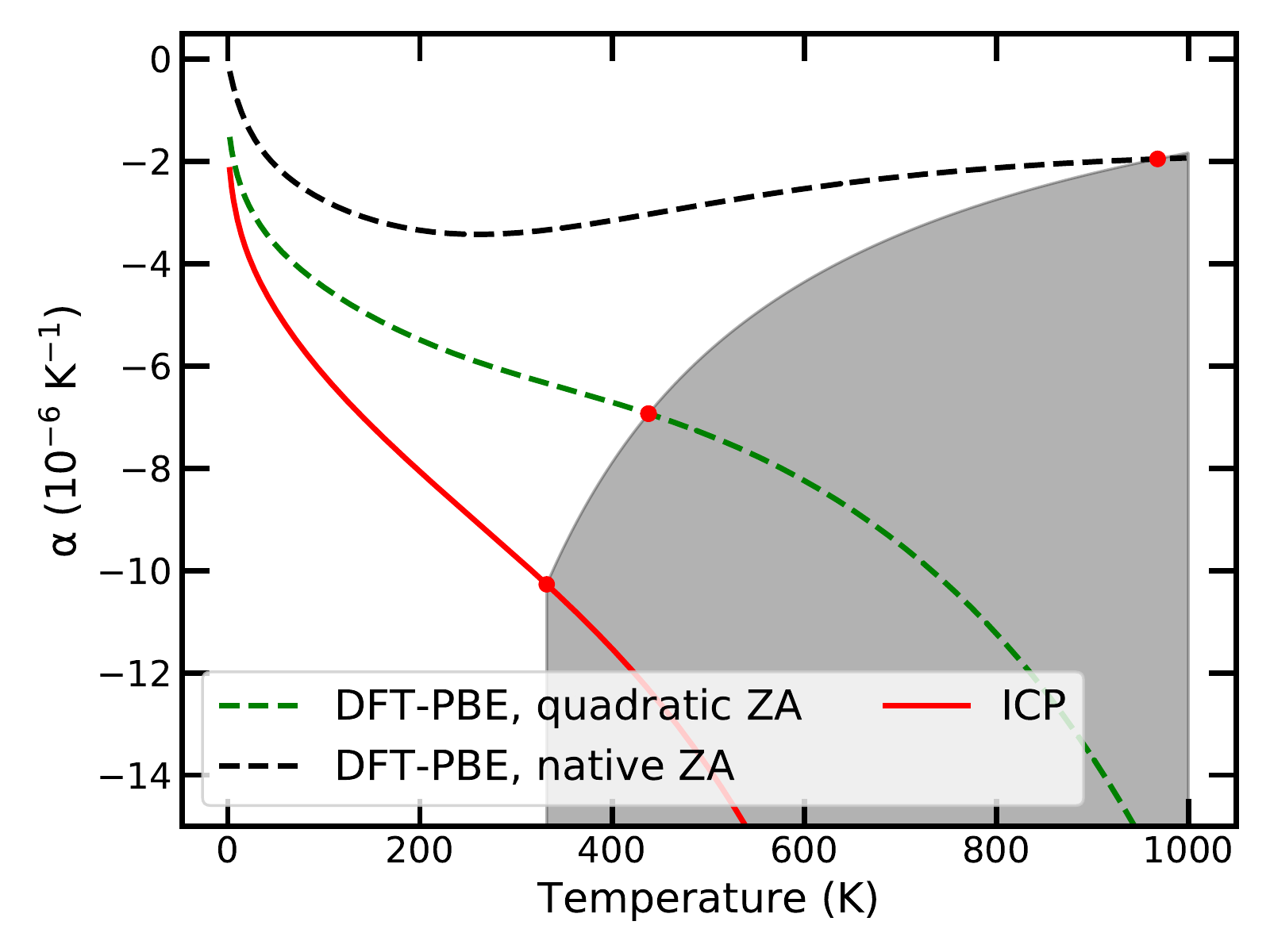}
\caption{ Linear thermal expansion coefficient $\mathrm{\alpha}$ for graphene.}
\label{alpha_fig}
\end{subfigure}
\begin{subfigure}{0.48\textwidth}
\includegraphics[scale = 0.5]{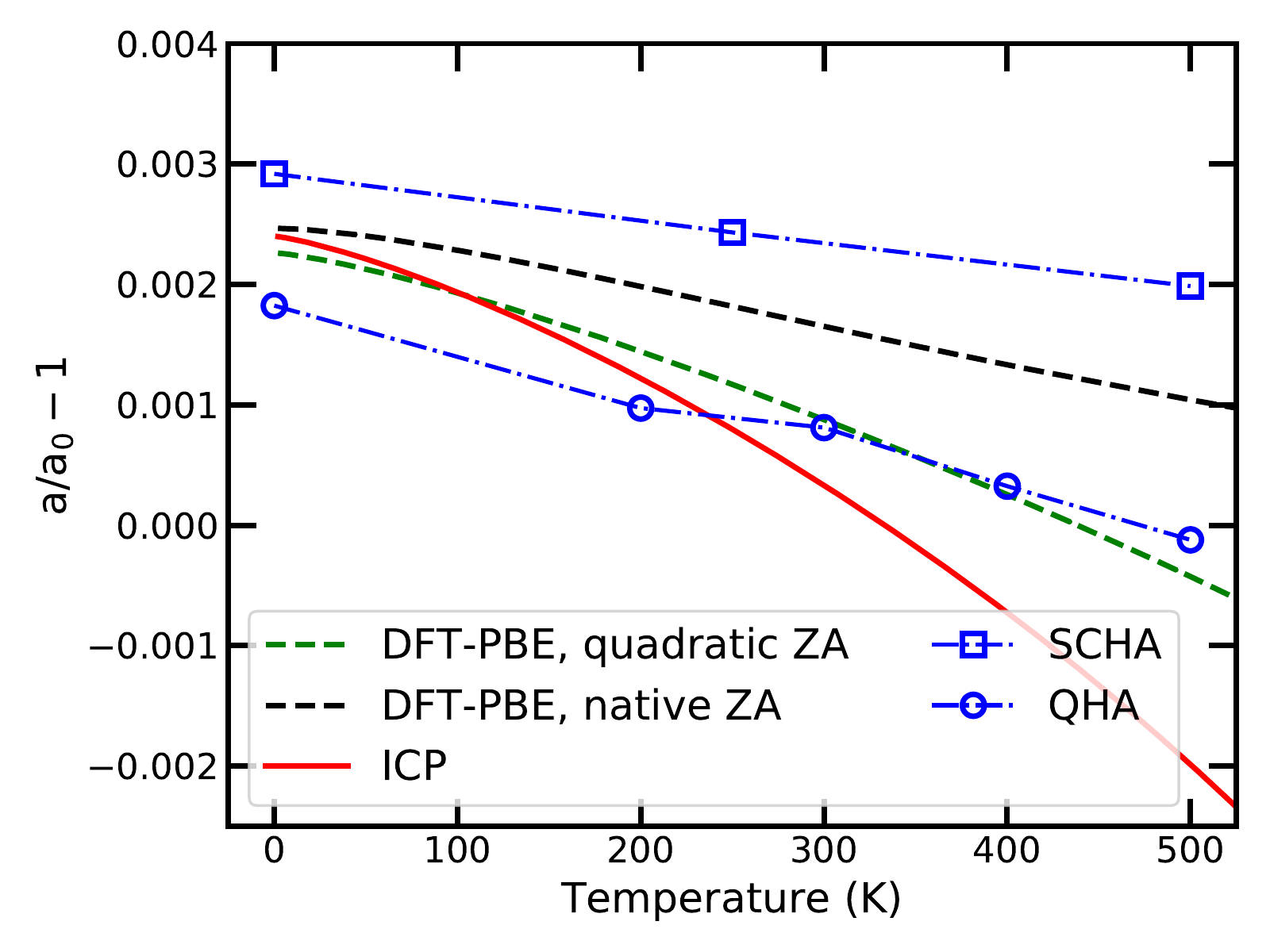}
\caption{Temperature dependence of the lattice parameter.}
\label{a_fig}
\end{subfigure}
\caption{Panel (a) shows the linear thermal expansion coefficients $\mathrm{\alpha}$ determined with the ICP (red solid line) and from first-principles, either without correction to the ZA mode (black dashed line) or enforcing  quadraticity close to $\mathrm{\Gamma}$, where it is affected by numerical noise (green dashed line). The shaded area indicates the range of temperatures in which the lattice parameter becomes smaller than $\mathrm{a_0}$, and the quasi harmonic approximation looses its validity. Panel (b) illustrates the behavior of the lattice parameter with the temperature. Square and round dots indicate, respectively, the values calculated by Ref. \cite{aseginolaza2020bending}  using the self-consistent harmonic approximation (SCHA) and the QHA with a Gaussian approximation potential (at 0 K, numerical noise bars the results to agree). }
\label{final_alpha_a}
\end{figure}
The linear thermal expansion coefficients calculated both from first-principles and with the ICP are presented in figure \ref{alpha_fig} . The difference between the first-principles curve obtained without applying corrections to the ZA mode (black dashed line) and the first-principles curve obtained by fitting a quadratic ZA (green dashed line) shows how dramatic is the effect of the quadraticity of the flexural phonon frequencies on the linear thermal expansion coefficient. In fact, the large negative Gruneisen coefficient for the ZA modes led Mounet and Marzari and to predict that graphene would contract with temperature \cite{PhysRevB.71.205214}, something that has been confirmed both in simulations \cite{PhysRevLett.102.046808} and experiments \cite{Yoon2011, LOPEZPOLIN2017670}. The first-principles results augmented with a quadratic fit have a trend that is similar to that of the ICP: both these curves start from around $\mathrm{-2\ 10^{-6}}$ and decrease monotonically until the lattice parameter becomes smaller than $\mathrm{a_0}$ and the QHA looses its validity. In fact if $\mathrm{a<a_0}$, the flexural phonon frequencies become imaginary, therefore the QHA predicts an imaginary vibrational energy for graphene. It is crucial to observe that, within the QHA, the lattice parameter at $\mathrm{T=0\ K}$ is larger than $\mathrm{a_0}$ (i.e. the one which is obtained by minimising the electronic energy of graphene without taking into account the vibrational contribution). The reason is that the phonon zero point energy causes a shift of the minumum of the free energy towards larger values of the lattice parameter, even at $\mathrm{0\ K}$. This shift is around 0.0025 $\mathrm{a_0}$ for both DFT-PBE and ICP. It is easy to obtain an analytic expression for this shift in the limit of small $\mathrm{\epsilon}$: the free energy of graphene at 0 K is given by
\begin{equation}
\mathrm{
    F(T=0\ K,a) = E(a) + \frac{1}{2}\sum_{\mathbf{q}s} \hbar \omega_{\mathbf{q}s}(a) \ ,
    }
\end{equation}
where E is the electronic energy, which has a minimum at $\mathrm{a=a_0}$. We can expand it around $\mathrm{a_0}$ as:
\begin{equation}
    \mathrm{
    E(a) \simeq \frac{1}{2}\lambda (a-a_0)^2\ ,
    }
\end{equation}
with 
$$\mathrm{\lambda = \frac{d^2E}{da^2}\ .}$$
The zero point energy contribution to the free energy can be expanded as well in the proximity of $\mathrm{a_0}$:
\begin{equation}
    \mathrm{
    \frac{1}{2}\sum_{\mathbf{q}s} \hbar \omega_{\mathbf{q}s}(a) \simeq 
    \frac{1}{2}\sum_{\mathbf{q}s} \hbar \omega_{\mathbf{q}s}(a_0)  
    - \sum_{\mathbf{q}s} \hbar\omega_{\mathbf{q}s}(a_0)\gamma_{\mathbf{q}s} a_0 (a-a_0)\ , 
    }
\end{equation}
where the definition of the Gr\"uneisen parameters in Eq. \ref{gru} has been used. The shift in the lattice parameter which minimises the free energy at 0 K is thus
\begin{equation}
    \mathrm{
    \frac{a}{a_0} - 1 =  \frac{1}{\lambda} \sum_{\mathbf{q}s} \hbar\omega_{\mathbf{q}s}(a_0)\gamma_{\mathbf{q}s}\ , 
    }
\end{equation}
which is around 0.0024 for both first-principles and ICP results, in very good agreement with the numerical minimization of the free energy. Without this shift, being the thermal expansion negative, the lattice parameter would be smaller than $\mathrm{a_0}$ for any $\mathrm{T>0\ K}$, leading to an ill-defined QHA.\\
The thermal expansion coefficient calculated through the ICP is slightly lower than the first-principles prediction; this is driven by the down-shift of the ICP Gr\"uneisen parameters for the modes LA and TA with respect to the first-principles values, as discussed above. The non-zero value of the lattice thermal expansion at 0 K is due to the singular behaviour of the ZA mode. 
It is interesting to compare the temperature dependence of the lattice parameter  calculated using the QHA with the prediction obtained under the framework of the self-consistent harmonic approximation (SCHA) \cite{PhysRevB.96.014111} by Aseginolaza et al. \cite{aseginolaza2020bending} (figure \ref{a_fig}). Both the ICP and the first-principles results underestimate the lattice parameter with respect to the SCHA prediction. This mismatch, which is even larger when considering the QHA results reported by Aseginolaza et al., is due in part to a quantitative inadequacy of the QHA; it is worth noting that the difference between the ICP and the first-principles results is much smaller than the error in the thermal expansion coefficient of the QHA, proving again that the ICP is able to reproduce first-principles results with great accuracy, and could be used for a full SCHA treatment, not suffering from the numerical noise of the machine-learned potential used in Ref. \cite{aseginolaza2020bending}.

\section{Thermal conductivity}
\noindent
Once harmonic and anharmonic force constants are correctly reproduced, the ICP can be tested on thermal transport properties, as e.g. obtained from the linearized Boltzmann transport equation (LBTE) \cite{peierls}. A first approximate estimate can be obtained with the single-mode approximation (SMA) \cite{PhysRevLett.106.045901}, corresponding to a kinetic picture of thermal transport in terms of single phonons: by neglecting the out-of-diagonal terms in the scattering matrix, the repopulation between different phonon modes is neglected.
An exact solution can be found by the variational method \cite{PhysRevB.75.035427}, or by iterative minimization \cite{OMINI1995101}. Although those approaches deliver the exact solution for the thermal conductivity, they do not provide any information of mean free paths for the carriers, and relaxation times, which are fundamental to characterise the thermal transport of a real material of finite size. In order to overcome this it is possible to express the exact solution of the LBTE by diagonalizing the full scattering matrix \cite{PhysRevX.6.041013}. 
This last approach leads to a picture of thermal transport where the carriers (relaxons) responsible for heat conduction are explicitly described.\

\begin{table}[h]
\begin{ruledtabular}
\begin{tabular}{lcr}
\multicolumn{1}{c}{} & ICP + cubic term & DFT-PBE\\
\hline
k SMA (W/mK) & 479 & 495\\
k variational (W/mK) & $\mathrm{3\ 650}$ & -$\,\,$ \\
k relaxons (W/mK) & $\mathrm{3\ 650}$ & $\mathrm{3\ 894}$ \\
Maximum MFP ($\mathrm{\mu}$m) & 24.9 & 23.2 \\
Maximum velocity (m/s) & 975 & 907 \\
Maximum relaxation time (ps) & 574 & 815 \\
\end{tabular}
\end{ruledtabular}
\caption{The values of the thermal conductivity k determined through the ICP are compared to those obtained from first-principles. The table also reports the maximum values of the mean free paths, relaxation times and velocities obtained with the two methods.}
\label{k_graphene}
\end{table}
The results obtained are summarized in table \ref{k_graphene}, including the thermal conductivity k obtained with the 3 approaches mentioned. The agreement between the SMA values for the thermal conductivity is remarkable, with an error of 3.2$\%$ with respect to the first-principles value. A very good result is obtained also when considering the exact thermal conductivity: the value calculated using the ICP is $\mathrm{3\ 650}\,\text{W}/\text{mK}$, 6.4$\%$ smaller with respect to the thermal conductivity calculated from first-principles ($\mathrm{3\ 894}\,\text{W}/\text{mK}$). 
Eventually, the ICP is able to reproduce not only the thermal conductivity value but also it correctly describes the properties of the heat carriers. 
\begin{figure*}[]
\centering
\begin{subfigure}{0.34\textwidth}
\includegraphics[scale = 0.18]{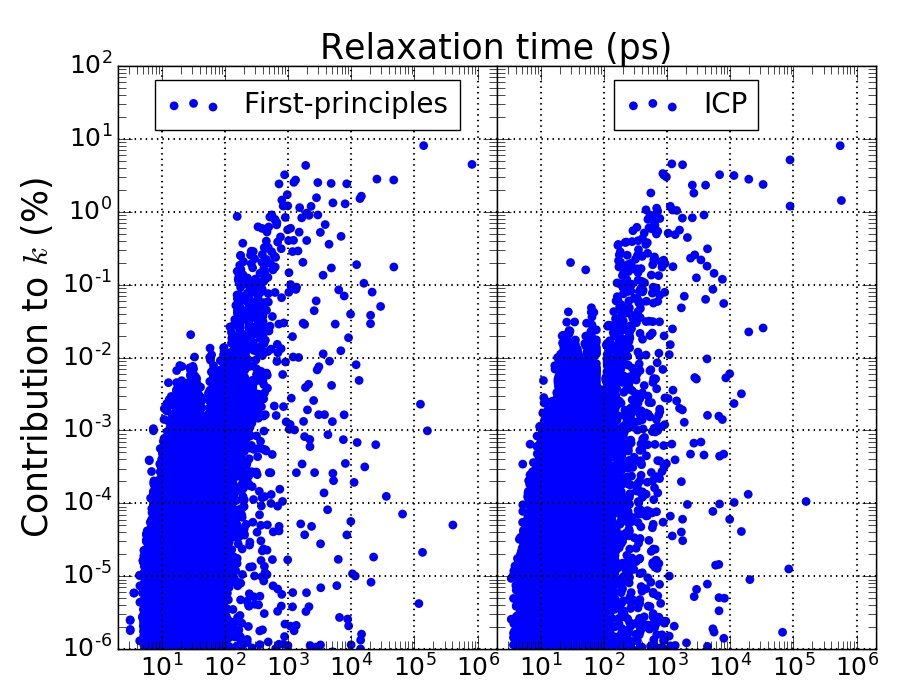}
\end{subfigure}
\begin{subfigure}{0.34\textwidth}
\includegraphics[scale = 0.18]{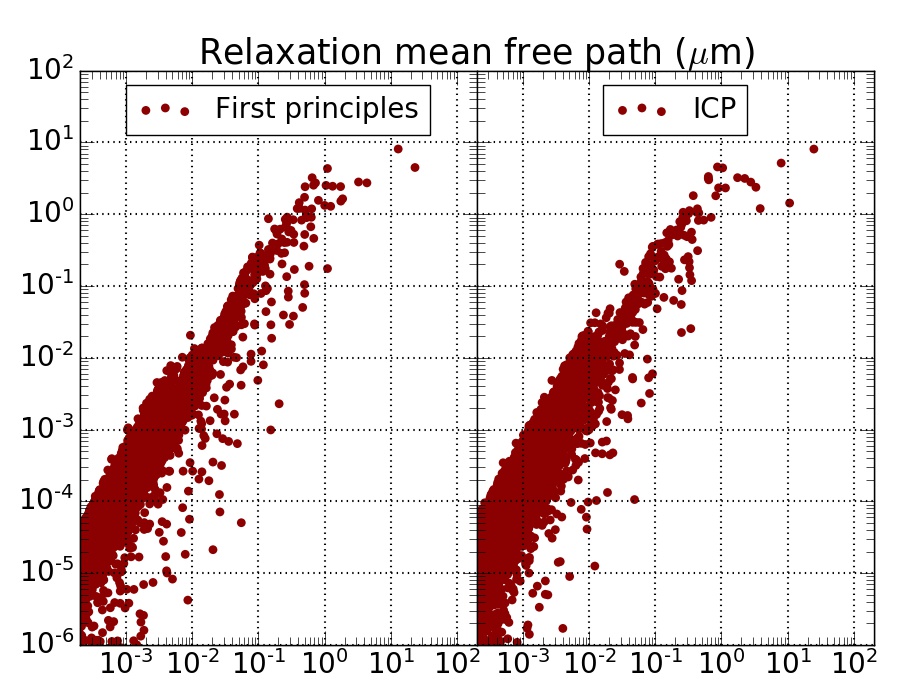}
\end{subfigure}
\begin{subfigure}{0.30\textwidth}
\includegraphics[scale = 0.18]{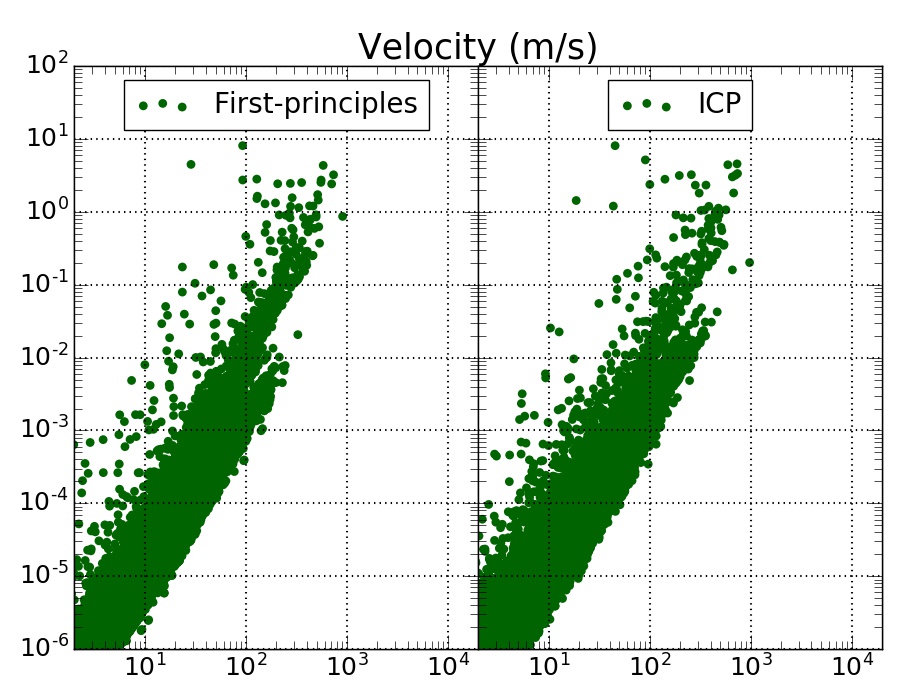}
\end{subfigure}
\caption{The contribution to thermal conductivity of the relaxons obtained with the ICP are compared to those calculated from first-principles \cite{PhysRevX.6.041013}, ordered according to their relaxation times (left), mean free paths (center) and velocities (right). }
\label{relaxons_graphene}
\end{figure*}
In figure \ref{relaxons_graphene} we show that the relaxons calculated from first-principles share many similarities with those calculated using the ICP: in both cases, only a limited number of relaxons contribute to the total thermal conductivity and, on average, corresponding relaxons have similar properties, with the maximum values for the relaxation times, the mean free paths and the velocities having the same order of magnitude (table \ref{k_graphene}). In particular, both simulations agree on the fact that the relaxons which contribute the most to thermal transport are those with the largest mean free path, which is around 25$\mathrm{\mu}$m. \\

\section{Conclusions}
\noindent
We have fitted the ICP introduced by Gartstein \cite{GARTSTEIN200483} onto first-principles calculations of the IFCs, and extended it with a single cubic stretching term in order to correctly reproduce both the harmonic and third-order anharmonic terms of the potential energy of honeycomb 2D materials and their nanotubes.
In particular, we focused on graphene, carbon nanotubes and 2D hexagonal boron nitride. 
The ICP has shown to give very good results for phonons in all cases considered, if compared to first-principles calculations; for graphene, it misses the Kohn anomalies at $\mathrm{\Gamma}$ and K, while for boron nitride it highlights how one should fit IFCs with the NACs removed at $\mathrm{q\neq0}$; these then should be added to the ICP at every q. This is broadly relevant for efforts where first-principles calculations are fitted with machine-learned potentials based on short-range descriptors: the fit should be performed on the analytic part of the IFCs, while the non-analytic effects should be modelled in reciprocal space, and added a posteriori.\\
Augmenting the Gartstein ICP with a simple cubic stretching term allows also to reproduce very well phonons in strained graphene for uniaxial and biaxial geometries, and the relative Gr\"uneisen parameters with overall close accuracy.
In addition,  the ICP is able to reproduce in full the quadraticity of the flexural modes, which greatly affect thermal transport and thermal expansion. The linear thermal expansion coefficient predicted by the ICP is close to that determined from first-principles, provided that, in the latter case, the quadraticity of the ZA mode close to the origine of the Brillouin zone is enforced with a parabolic fit; in fact, the thermal expansion coefficients are dominated by the contribution of the ZA Gr\"uneisen parameters, which diverge as $\mathrm{q^{-2}}$ in the long-wavelength limit. Linearization of the ZA frequencies close to the $\mathrm{\Gamma}$ point due to numerical noise in first-principles simulations prevents the ZA Gr\"uneisen parameter from diverging, leading to an important underestimation of the thermal expansion. Such a quadratic corrections of the ZA frequencies is not required for the ICP, due to the parabolic shape of the ZA mode also for very small momenta.\\
We tested the reliability of the ICP in calculating graphene's thermal conductivity, either in the single-mode approximation or in the exact solution. For the former case the error is almost negligible, while for the latter one it is around 6.4$\%$. The ICP also reproduces correctly mean free paths, velocities and lifetimes for relaxons, when compared to those obtained from the first-principles. We thus conclude that such anharmonic ICP, fitted on first-principles data, is a valuable tool to perform thermomechanical simulations on honeycomb materials, since it provides good to excellent accuracy, and noiseless results, at a computational cost that is negligible, particularly if compared to the one of first-principles calculations or even machine-learned potentials.\\
Regarding the future perspectives, the addition of additional cubic terms would lead to an even better
agreement with first-principles thermal conductivity - that nevertheless
already now seems excellent. The present analytic expression for the potential, or some
small modifications, could provide very good results for other materials
with a buckled hexagonal structure, such as silicene, germanene and
stanene (group IV) or phosphorene, arsenene and antimonene (group V);
graphene oxide or oxygenated graphene would also be ready for studies
and development. But probably most importantly the current potential
could be used to studies of the thermomechanical properties of graphene
and boron nitride membranes at time scales and length scales that are
relevant to experimental and technological applications \cite{Cohen-Tanugi2012,Huang2018,Ramanathan2018}, and we see this
as one of the most exciting avenues forward.

\section{Methods}\label{methods}
\noindent

First-principles calculations have been performed using the open-source Quantum ESPRESSO distribution \cite{Giannozzi_2009}, using the PBE exchange-correlation functional, and pseudopotentials for carbon \cite{DALCORSO2014337}, boron \cite{GARRITY2014446} and nitrogen \cite{VANSETTEN201839} as suggested by the SSSP Precision library \cite{Prandini2018} version 1.1. The planewave cutoff used for both graphene and carbon nanotubes is $\text{80}\ \text{Ry}$, while for hexagonal boron nitride is $\text{100}\ \text{Ry}$. The charge density cutoff used for all the materials is 12 times larger. For the self-consistent calculations the Brillouin zone has been sampled with a 12$\times$12$\times$1 unshifted grid for graphene and hexagonal boron nitride, and a 1$\times$1$\times$12 unshifted grid for the nanotubes. For phonon calculations, the q-point grid used is 10$\times$10$\times$1 for graphene, 12$\times$12$\times$1 for hexagonal boron nitride and 1$\times$1$\times$12 for carbon nanotubes.\\
To converge the linear thermal expansion coefficients, the BZ has been discretized with a 128x128 q-points grid, both for the ICP and for the first-principles calculations.\\
The third derivatives are computed on a 4x4 supercell, using the open source software Phono3py \cite{phono3py}.\\
In order to compare the ICP results  with those determined from first-principles, we calculate the thermal conductivity of graphene using the same parameters as Ref. \cite{doi:10.1021/nl502059f}: the Dirac's delta in the scattering expression is broadened with a Gaussian smearing of $\text{10}\ \text{cm}^{-1}$, the Brillouin zone is discretized with a 128$\times$128$\times$1 q-point phonon grid, and the equivalent thickness used to compare the 2D thermal conductivity of graphene to that of 3D materials is taken to be the experimental inter-layer distance of graphite (3.32 \AA) \cite{PhysRevLett.118.135901}. The scattering matrix is built considering three-phonon scattering due to the anharmonicities and two-phonons events linked to the presence of carbon isotopes at natural abundance. \\

\section{Acknowledgments}\noindent
This project has received funding from the European union's Horizon 2020 research and innovation program under the Marie Skłodowska-Curie grant agreement \textnumero$\ $754354, and has been in part supported by NCCR MARVEL. Simulation time was 
awarded by PRACE on Marconi at Cineca, Italy (project id. 2016163963).
We thank also Andrea Cepellotti and Michele Simoncelli for the useful discussions.




\bibliography{main}

\end{document}